\documentclass[aps,prd,10pt, tightenlines, twoside, secnumarabic, 
superscriptaddress,showpacs, preprintnumbers, showpacs, nofootinbib, 
twocolumn]{revtex4-1}

\pdfoutput=1
\usepackage{amsthm}
\usepackage{mathtools}
\usepackage{amsfonts}
\usepackage[left=1.5cm,right=1.5cm,top=2cm,bottom=2cm]{geometry}
\usepackage{amssymb}
\usepackage{graphicx}
\usepackage{multirow}
\usepackage{color}
\newcommand{\beq}{\begin{eqnarray}}
\newcommand{\eeq}{\end{eqnarray}}

\newcommand{\bmp}{\noindent\begin{minipage}{16cm}}
\newcommand{\emp}{\end{minipage}\vskip 7mm} % 7mm untightened

\usepackage{dcolumn}% Align table columns on decimal point
\usepackage{bm}% bold math
\usepackage{bbm}
\usepackage{subfigure}
\usepackage{slashed} 
\usepackage{mathrsfs}

\theoremstyle{definition}

\theoremstyle{plain}

\usepackage{epsfig}

\usepackage[ margin=5pt, font=normalsize,labelfont=bf,justification=raggedright]{caption}

\usepackage{hyperref}
\definecolor{rossoCP3}{cmyk}{0,.88,.77,.40}
\definecolor{verdeCP3}{rgb}{0.09765625, 0.57421875, 0.1015625}
\definecolor{bluCP3}{rgb}{0, 0.23, 0.67}
\hypersetup{colorlinks, bookmarksopen, bookmarksnumbered,citecolor=bluCP3, linkcolor=bluCP3, pdfstartview=FitH, urlcolor=bluCP3}
\def\lsim{\mathrel{\rlap{\lower4pt\hbox{\hskip1pt$\sim$}}
    \raise1pt\hbox{$<$}}}                % less than or approx. symbol
\def\gsim{\mathrel{\rlap{\lower4pt\hbox{\hskip1pt$\sim$}}
    \raise1pt\hbox{$>$}}}                % greater than or approx. symbol

\newcommand{\bea}{\begin{eqnarray}}
\newcommand{\eea}{\end{eqnarray}}

\newcommand{\ba}{\begin{eqnarray}}
\newcommand{\ea}{\end{eqnarray}}

\newcommand{\be}{\begin{eqnarray}}
\newcommand{\ee}{\end{eqnarray}}

\begin{document}
\font\secret=cmr10 at 0.8pt
%%%%%%%%%%%%%%%%%%%%%%%%%%%%%%%%%%%%%%%%%%%%%%%%%%%%%%%%%%%%%%%%%%%%%%%%%%%
\title{ ~~\\  Asymmetric Dark Matter Stars}
%%%%%%%%%
% 

%$^{\color{rossoCP3}{\varheartsuit}}$}
\author{Chris Kouvaris}
\email{kouvaris@cp3.dias.sdu.dk}
\author{Niklas Gr\o nlund Nielsen}
\email{ngnielsen@cp3.dias.sdu.dk}

\affiliation{CP$^{3}$-Origins \& Danish Institute for Advanced Study  DIAS, University of Southern Denmark, Campusvej 55, DK-5230 Odense M, Denmark}

%%%%%
%%%%%%%%%%%%%%%%%%%%%%%%%%%%%%%%%%%%%%%%%%%%%%%%%%%%%%%%%%%%%%%%%%%%%%%%%%%

\begin{abstract}
We study the possibility of asymmetric dark matter with self-interactions forming compact stable objects. We solve the Tolman-Oppenheimer-Volkoff equation and 
find the mass-radius relation of such ``dark stars", their density profile and their Chandrasekhar mass limit. We consider fermionic asymmetric dark matter with Yukawa-type self-interactions appropriate for solving the well known problems of the collisionless dark matter paradigm. We find that in several cases the relativistic 
effects are significant.
% \textcolor{red}{Perhaps replace last sentence with: We find that relativistic effects may be important if the Yukawa interactions are repulsive, and that the limiting factor with attractive interactions is that the pressure %vanishes as the Fermi momentum increases.}
\\[.1cm]
{\footnotesize  \it Preprint: CP3-Origins-2015-027 DNRF90, DIAS-2015-27}
 \end{abstract}

\maketitle
%\thispagestyle{fancy}
%\tableofcontents
%\newpage

\section{Introduction}

Lately, an emerging amount of issues indicates that the Collisionless Cold Dark Matter (CCDM) paradigm is at odds with astrophysical observations. The first and most well known issue is related to the fact that dwarf galaxies have a flat density core~\cite{Moore:1994yx,Flores:1994gz}. Dwarf galaxies are dominated by dark matter (DM) and the flatness of the density profile in the core of the galaxy is in contradiction with the cuspy profiles predicted by numerical simulations of CCDM~\cite{Navarro:1996gj}. Numerical simulations of CCDM predict also a larger number of satellite galaxies in the Milky Way than what has been observed so far~\cite{Klypin:1999uc,Moore:1999nt,Kauffmann:1993gv}. Although our galaxy might simply be  a statistical fluctuation~\cite{Liu:2010tn,Tollerud:2011wt,Strigari:2011ps} there might be dim galaxies yet to be observed, it is possible that DM is simply not collisionless. Furthermore, another related issue  is the so-called  ``too big to fail" problem~\cite{BoylanKolchin:2011de}, i.e. CCDM numerical simulations predict massive dwarf galaxies that are too big to not have visible stars and thus  to not be observed.  Although some of the aforementioned issues can be resolved upon assuming the existence of baryonic-DM interactions~\cite{Oh:2010mc,Brook:2011nz,Pontzen:2011ty,Governato:2012fa}, another probably more attractive and natural possibility is the existence of DM self-interactions. Clearly such interactions would flatten out cuspy dwarf galaxy cores and they could possibly also resolve  the satellite galaxies issues~\cite{Vogelsberger:2012ku,Rocha:2012jg,Zavala:2012us,Peter:2012jh}.

DM self-interactions have already been  proposed and studied in different contexts~\cite{Spergel:1999mh,Wandelt:2000ad,Faraggi:2000pv,Mohapatra:2001sx,Kusenko:2001vu,Loeb:2010gj,Kouvaris:2011gb,
Rocha:2012jg,Peter:2012jh,Vogelsberger:2012sa,Zavala:2012us,Tulin:2013teo,Kaplinghat:2013xca,Kaplinghat:2013yxa,Cline:2013pca,
Cline:2013zca,Petraki:2014uza,Buckley:2014hja,Boddy:2014yra,Schutz:2014nka}. DM numerical simulations including self-interactions favor a DM-DM cross section per DM mass between $0.1-10~\text{cm}^2/\text{g}$. Within this range, DM self-interactions can solve the cusp vs core problem of dwarf galaxies as well as the ``too big to fail" one. However, DM self-interactions cannot be arbitrarily strong. There are several constraints imposed on them. First of all one should make sure that DM-DM interactions are not sufficiently strong to destroy the ellipticity of spiral galaxies~\cite{Feng:2009mn,Feng:2009hw} or dissociate the subclusters of the bullet cluster~\cite{Markevitch:2003at}. In addition, fermionic asymmetric DM with attractive Yukawa-type self-interactions can lead in some cases to formation of destructive black holes in the interior of old neutron stars, thus imposing extra constraints~\cite{Kouvaris:2011gb}. Further constraints are imposed in the case where the mediator of the DM-DM force couples to the Standard Model. The mediator $\phi$ can simply couple to the Standard Model via e.g. a Higgs portal~\cite{Burgess:2000yq,Patt:2006fw,Andreas:2008xy,Andreas:2010dz,Djouadi:2011aa,Pospelov:2011yp,Greljo:2013wja,Bhattacherjee:2013jca}. In such a case one should make sure that $\phi$ decays before the start of the Big Bang Nucleosynthesis. The fact that a minimum strength between baryons and $\phi$ is required for not spoiling the BBN predictions can lead to significant rates of DM collisions in underground detectors that can exclude such models~\cite{Kaplinghat:2013yxa}. However constraints like these can  be evaded if e.g. $\phi$ couples also to sterile or active neutrinos~\cite{Kouvaris:2014uoa}. One should emphasize that the above constraints are model dependent and therefore although there is no clear universal region where DM self-interactions are allowed, the region $0.1-1~\text{cm}^2/\text{g}$ which accommodate the resolution of the dwarf galaxies problems is roughly speaking constraint free.  

If DM experiences self-interacting forces, it is possible to imagine that star-like compact objects can be formed. Whatever the mechanisms of forming such objects, one should make sure that they do not violate the limits imposed by the MACHO~\cite{Alcock:2000ph} and EROS~\cite{Tisserand:2006zx} experiments. Based on the microlensing technique, these experiments claimed that less than $20\%$ of DM can be in the form of compact objects between the mass range $10^{-7}M_{\odot}\lesssim  M \lesssim 10 M_{\odot}$, where $M_{\odot}$ is the solar mass. The possibility of stars made of DM has been studied before in the context of annihilating DM forming dark stars in the early universe~\cite{Spolyar:2007qv,Freese:2008hb,Freese:2008wh}. It has been also studied in the context of hybrid compact stars made of baryonic and DM~\cite{Leung:2011zz,Leung:2013pra}. In the latter case, neutron stars and white dwarfs include a significant amount of DM in their interior modifying thus the equation of state of the star. Furthermore the possibility of black hole formation from strongly self-interacting components of DM was studied recently in~\cite{Pollack:2014rja}.

In this paper we examine the possibility that asymmetric DM with self-interactions appropriate for solving the core vs cusp problem, the ``satellite problem" and the ``too big to fail problem" forms star-like compact objects. Asymmetric DM~\cite{Nussinov:1985xr,Barr:1990ca,Gudnason:2006yj,Foadi:2008qv,Dietrich:2006cm,Sannino:2009za,Ryttov:2008xe,Sannino:2008nv,Kaplan:2009ag,Frandsen:2009mi,MarchRussell:2011fi,
Frandsen:2011cg,Gao:2011ka,Arina:2011cu,Buckley:2011ye,Lewis:2011zb,Davoudiasl:2011fj,Graesser:2011wi,Bell:2011tn,Cheung:2011if}
has become an attractive alternative to thermally produced DM not only because it can relate theories beyond the Standard Model to DM, but because it can also provide a link between baryogenesis and dark-genesis. For recent reviews on asymmetric DM see~\cite{Petraki:2013wwa,Zurek:2013wia}.
We are going to assume that the self-interactions are Yukawa-type and can  be either  attractive (mediated by  a  scalar $\phi$) or repulsive (mediated by a vector boson $\phi_{\mu}$). Upon these assumptions, we study the stability of dark stars formed by asymmetric fermionic DM. We solve the Tolman-Oppenheimer-Volkoff equation and study the hydrostatic equilibrium of these compact objects. We find their density profile, the mass vs radius relation as well as the Chandrasekhar mass, i.e. the maximum mass where these objects are stable.

The paper is organised as follows: In section II we present the equation of state for DM with self-interactions. In section III we show the relevant parameter phase space of self-interactions that solve the problems of CCDM we have mentioned earlier. In section IV we present the equations for the hydrostatic equilibrium of the dark stars. In section V we use a simplified Newtonian analysis to get a first understanding of the problem, while we present the full relativistic results in section VI. We conclude in section VII.

Throughout this paper we use natural units $\hbar = c = k_\text{B}=1$, and define the Planck mass as $M_\text{P} = G^{-1/2}$.

\section{Equation of State}
As we mentioned we assume that DM is  of asymmetric type and fermionic. We also assume that DM self-interactions are mediated by $\phi$ via a Yukawa coupling of the form $g\phi \bar{\chi} \chi$ (in case of attractive interactions) where $\chi$ is the DM particle and $g$ is the Yukawa coupling constant or $g \phi_\mu \bar{\chi} \gamma^\mu \chi$ (in case of repulsive interactions).
We wish to obtain the equation of state of DM under the aforementioned assumptions. The energy density of DM particles $\rho$ consists of two components 
\begin{equation}
\rho=\rho_{\text{kin}}+\rho_{\text{Y}},
\end{equation}
where   $\rho_{\text{kin}}$, and $\rho_{\text{Y}}$ are the kinetic energy density, and Yukawa potential energy respectively. The pressure of the system is~\cite{Shapiro:1983du} 
\begin{equation}
P= n^2\frac{d}{dn}\left(\frac{\rho}{n} \right),
\end{equation}
where $n$ is the number density of DM particles. The pressure related to the kinetic energy $P_{\text{kin}}$  in the non-relativistic ($ p\ll m_\chi$) or relativistic ($ p \gg m_\chi$) limits takes  the simple form of a polytrope $P_{\text{kin}}=K\rho_{\text{kin}}^{\Gamma}$, where $\Gamma$ equals $5/3$ or $4/3$ for the non-relativistic and relativistic case respectively. We choose  to work with the full relativistic dispersion $E_{\text{kin}} = \sqrt{p^2+m_\chi^2}$. If we assume that the temperature of DM particles is much smaller than their Fermi energy, i.e. we effectively take the limit $T=0$, the number density, kinetic energy density and the corresponding pressure are given by 
\begin{equation}
n = \frac{g_s}{(2\pi)^3} \int_0^{p_{F}} 4\pi p^2 dp =  \frac{g_s m_\chi^3}{6\pi^2} x^3,
\end{equation}

\begin{equation}
\rho_{\text{kin}} = \frac{g_s}{(2\pi)^3}\int_0^{p_F} E(p) \;4\pi p^2 dp=\frac{g_s}{2}m_\chi^4 \xi(x), \label{eq:kinetic density}
\end{equation}
\begin{equation}
P_{\text{kin}} = \frac{1}{3}\frac{g_s}{(2\pi)^3}\int_0^{p_F} \frac{p^2}{E(p)} \;4\pi p^2 dp= \frac{g_s}{2}m_\chi^4\psi(x),
\end{equation}
where the functions $\xi$ and $\psi$ are defined as \\
\begin{equation}
 \xi(x) = \frac{1}{8\pi^2} \left\{ x\sqrt{1+x^2}\,(1+2x^2)- \ln \left[x+\sqrt{1+x^2}\,
 \right]\right\},
\end{equation} 
\begin{equation}
\psi(x) = \frac{1}{8\pi^2} \left\{ x\sqrt{1+x^2}\,(2x^2/3-1)+ \ln \left[x+\sqrt{1+x^2}\,
 \right]\right\}. \label{7}
\end{equation}
 $x = p_F/m_\chi$ is a measure of how relativistic the particles are, and $g_s=2 s+1$ is the spin multiplicity.

The Yukawa potential between two particles is
\begin{equation}
V_{ij}= \pm \alpha \frac{e^{-\mu r_{ij}}}{r_{ij}},
\label{eq:Yukwabetween2particles}
\end{equation}
where $\mu$ is the mass of the mediator, $r_{ij}$ is the separation between the particles and $\alpha = g^2/4\pi$ is the coupling to the dark mediator. In order to find the Yukawa potential energy of the entire system we have in principle to sum over all pairs of DM particles, which we approximate as an integration over volume elements.
\begin{equation}
E_{\text{Y}} = \frac{1}{2} \sum_{i\neq j} V_{ij} = \pm \frac{1}{2} n^2 \alpha \int \int \frac{e^{-\mu r_{ij}}}{r_{ij}} dV_i dV_j.
\end{equation}
 In the case where the radius of the star satisfies $R \gg 1/\mu$ (i.e. the potential is short range), it is a reasonable approximation to integrate the volume up to infinity (instead of the volume of the star).  This leads to the following Yukawa energy density
\begin{equation}
\rho_{\text{Y}} = \pm \frac{2 \pi \alpha n^2}{\mu^2} = \pm \frac{\alpha g_s^2}{18 \pi^3} \frac{m_\chi^6}{\mu^2} x^6.
\end{equation}
This estimate for the Yukawa energy density gives us the final expressions for $P$ and $\rho$:

\begin{align}
P=\frac{g_s}{2}m_\chi^4\psi(x) \pm \frac{\alpha g_s^2}{18 \pi^3} \frac{m_\chi^6}{\mu^2} x^6, \label{eqstate1}\\
\rho = \frac{g_s}{2}m_\chi^4\xi(x) \pm \frac{\alpha g_s^2}{18 \pi^3} \frac{m_\chi^6}{\mu^2} x^6 \label{eqstate2}.
\end{align}
Since neither equation can be inverted analytically, we must work with two equations of state, and have an implicit relation between $P$ and $\rho$.
  DM particles in the attractive scenario (corresponding to the minus sign in the Yukawa contribution) cannot become arbitrarily relativistic, since the pressure and density must be positive. The positiveness of pressure and density give an upper bound on $x$.

\section{Parameter Space of Self-Interactions}
As we mentioned earlier, DM self-interactions within the range  $\sigma / m_\chi =  0.1-10$ cm$^2$/g can solve the problematic issues of CCDM, while astrophysical constraints limit these interactions between   $\sigma / m_\chi =  0.1-1$ cm$^2$/g. Following~\cite{Tulin:2013teo}, in order to determine the parameter space of DM and mediator masses (for a given coupling) that lies in the aforementioned range, we introduce the transfer cross section $\sigma_T = \int d\Omega (1-\cos\theta) d\sigma / d\Omega$. We use a typical value of $v_0 = 10$ km/s for the average velocity of DM in a dwarf galaxy, and we estimate the velocity averaged cross section as
\begin{equation}
\sigma = \int d^3 v \frac{e^{- (v/v_0)^2/2}}{(2\pi v_0^2)^{3/2}} \sigma_T(v).
\end{equation}

\begin{figure*}
\begin{tabular}{cc}
  \includegraphics[width=.35\textwidth]{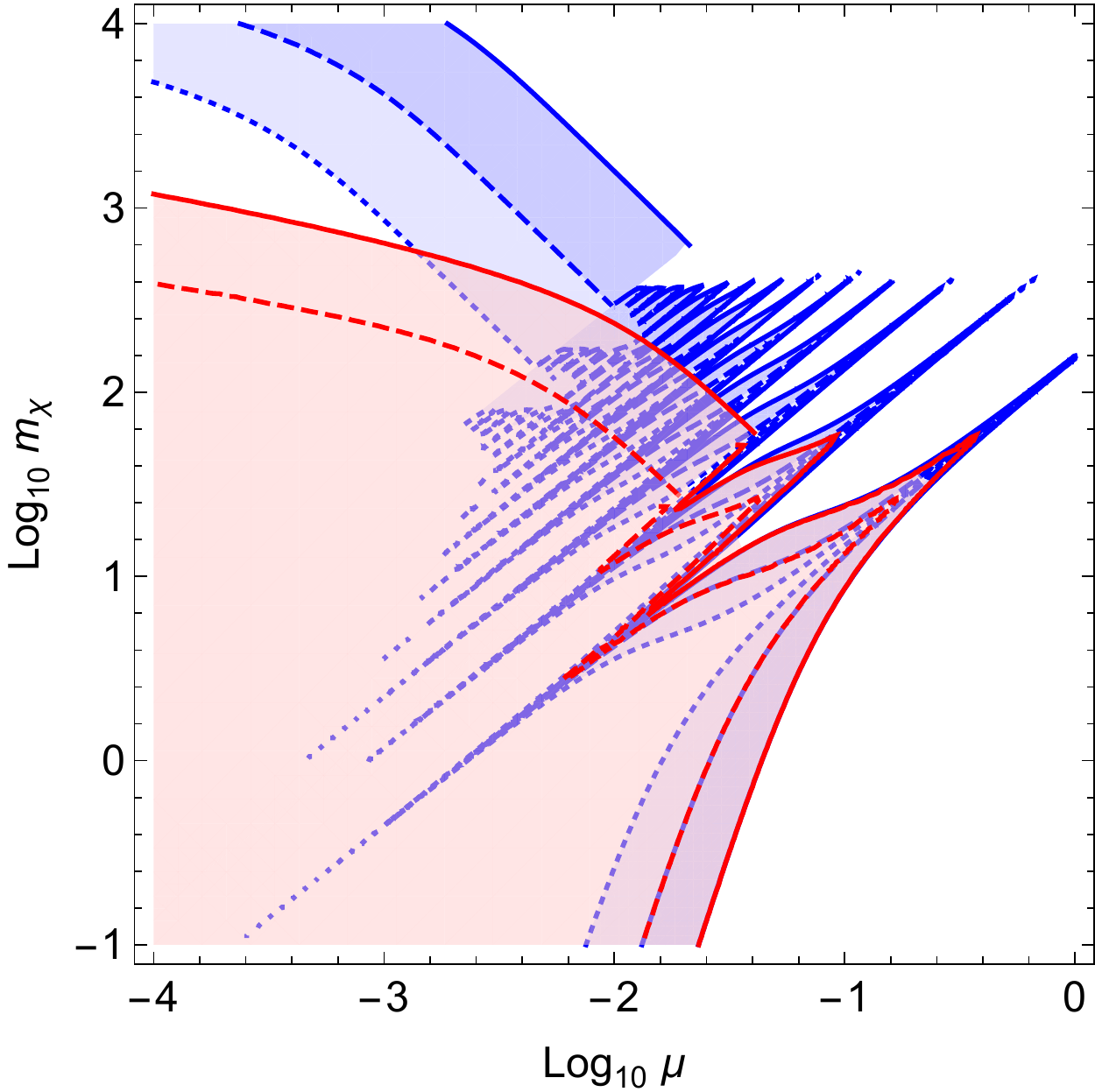} &   \includegraphics[width=.35\textwidth]{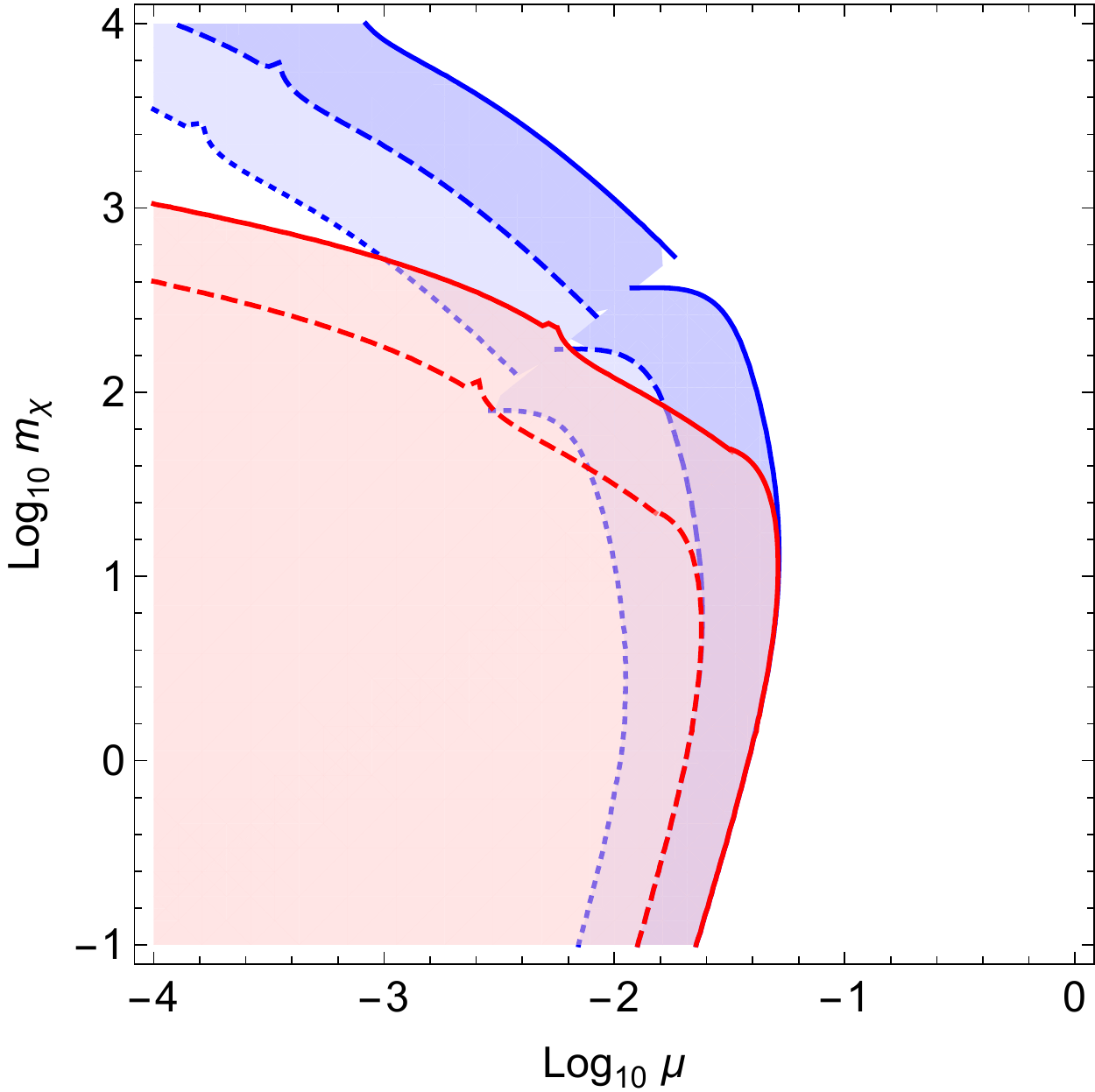} \\
(a) $\alpha = 10^{-2}$ attractive & (b) $\alpha = 10^{-2}$ repulsive \\[6pt]
  \includegraphics[width=.35\textwidth]{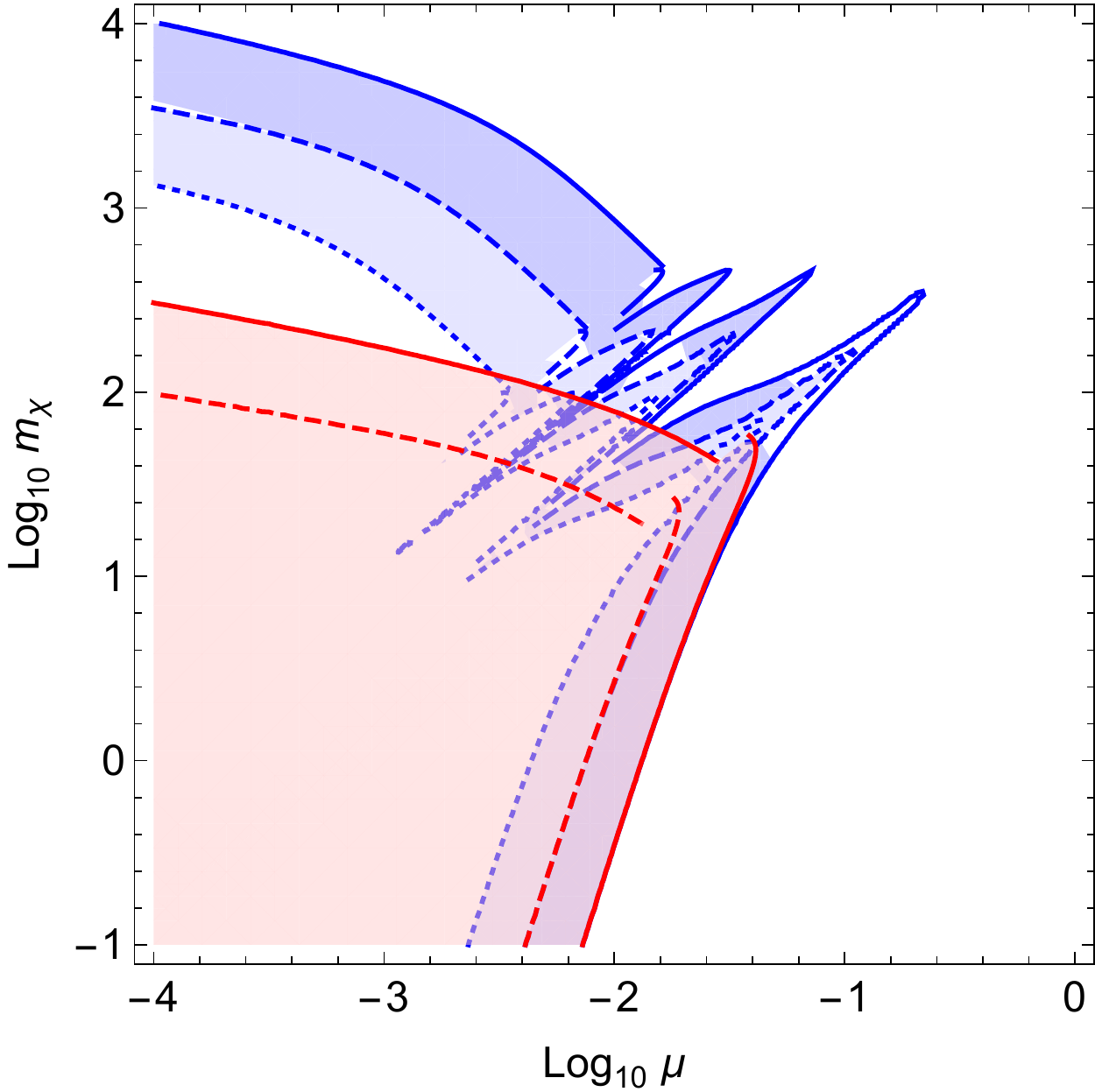} &   \includegraphics[width=.35\textwidth]{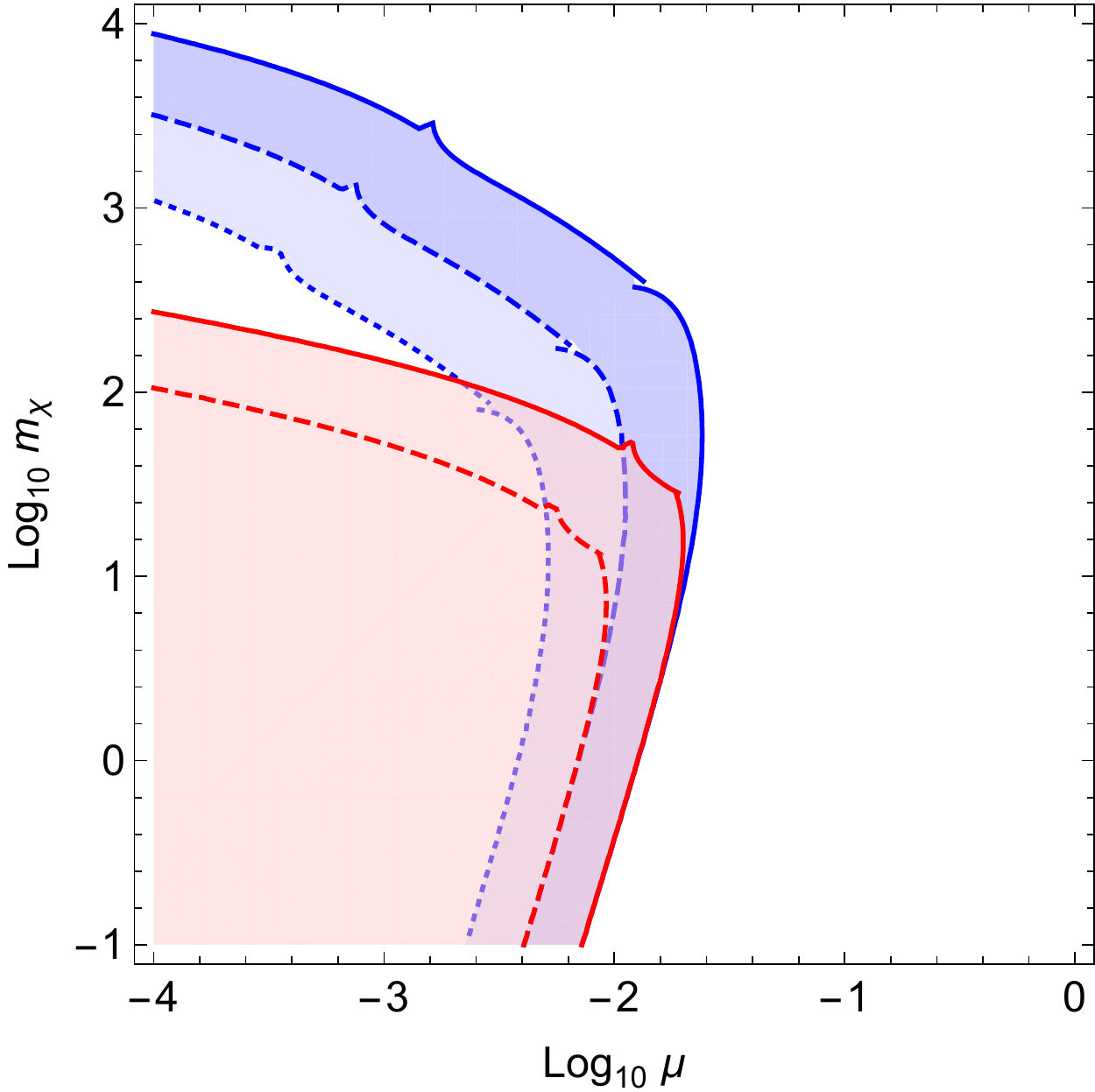} \\
(c) $\alpha = 10^{-3}$ attractive & (d) $\alpha = 10^{-3}$ repulsive \\[6pt]
  \includegraphics[width=.35\textwidth]{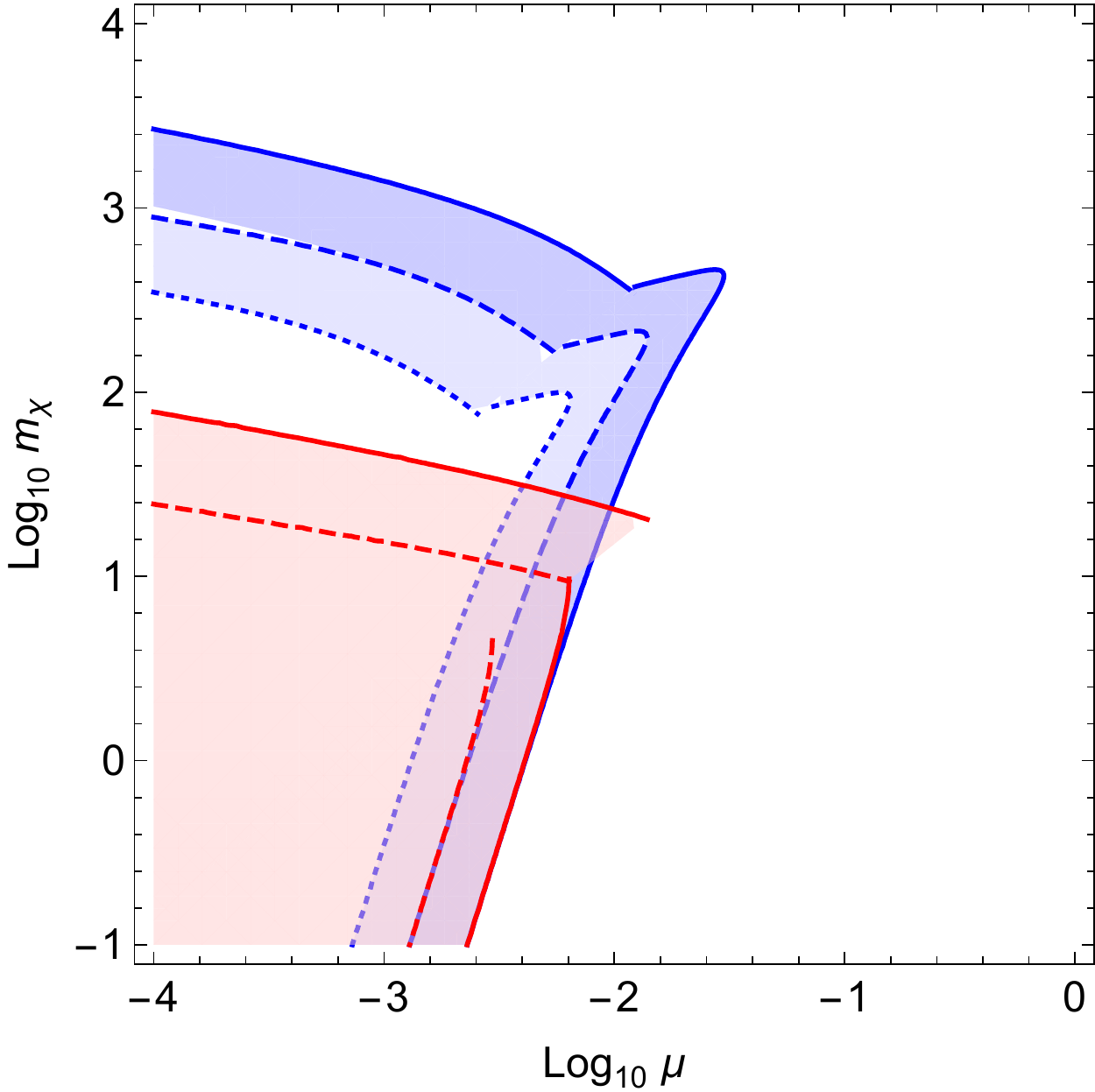} &   \includegraphics[width=.35\textwidth]{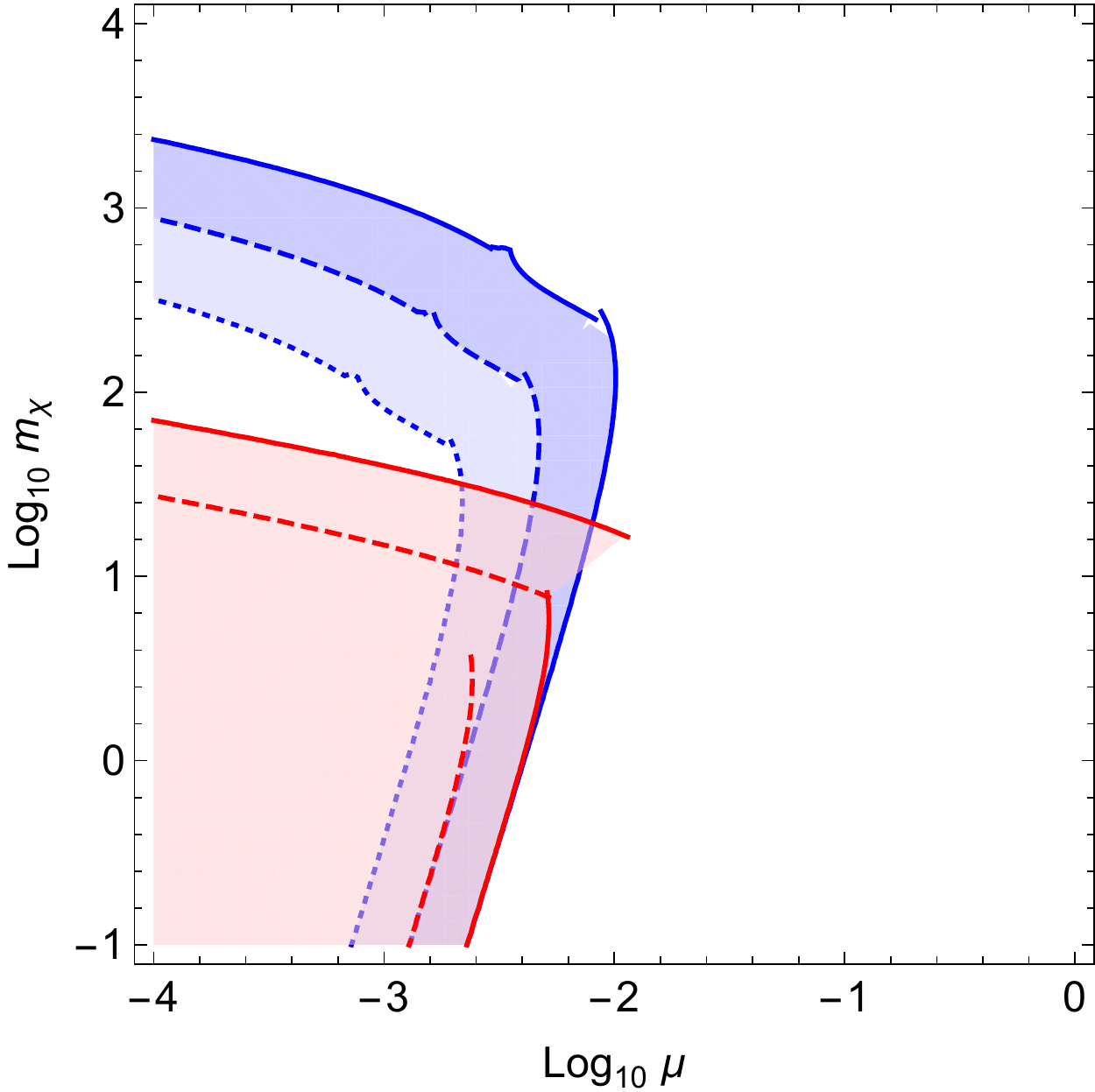} \\
(e) $\alpha = 10^{-4}$ attractive & (f) $\alpha = 10^{-4}$ repulsive
\end{tabular}
\caption{The parameter phase space for DM and mediator masses (measured in GeV) that can solve the problems of  CCDM. Left panels correspond to attractive interactions while the right panels to repulsive ones. We show three different values of the Yukawa strength $\alpha$. See text for details.}
\label{fig:parameter}
\end{figure*}

The transfer cross section for attractive Yukawa interactions in the classical limit $m_\chi v/\mu \gg 1$ is~\cite{Feng:2009hw,Finkbeiner:2010sm} 
\begin{equation}
\sigma_T= 
\begin{dcases}
 \frac{4\pi}{\mu^2} \beta^2 \log \left( 1 + \beta^{-1}\right)& \text{if } \; \; \beta \lesssim  10^{-1}\\
 \frac{8\pi}{\mu^2} \beta^2 / \left( 1 + 1.5 \beta^{-1.65}\right)& \text{if } \; \;  10^{-1} \lesssim \beta \lesssim  10^{3}\\
  \frac{\pi}{\mu^2}  \left( \log \beta + 1 - \tfrac{1}{2} \log ^{-1} \beta\right)^2& \text{if } \; \; \beta \gtrsim  10^{3}.\\
\end{dcases}
\end{equation}
The corresponding cross section for repulsive Yukawa interactions interactions reads
\begin{equation}
\sigma_T= 
\begin{dcases}
 \frac{2\pi}{\mu^2} \beta^2 \log \left(1+\beta^{-2} \right),& \text{if } \; \; \beta \lesssim  1\\
 \frac{\pi}{\mu^2}\left( \log 2\beta - \log \log2\beta\right)^2,& \text{if } \; \;  \beta \gtrsim  1\\
\end{dcases}
\end{equation}
where $\beta = 2\alpha \mu /(m_\chi v^2)$. For  the opposite limit $m_\chi v/\mu \leq 1$, we follow~\cite{Tulin:2013teo} and approximate the Yukawa  by a Hulth\'{e}n potential  with proper parameter choices. In the region $m_\chi v/\mu \sim 1$ we interpolate between the two regimes. In Fig.~\ref{fig:parameter} we show the allowed parameter space for DM and mediator mass (in GeV) in the case of repulsive and attractive interactions for three distinct values  $\alpha=10^{-2},~10^{-3},~10^{-4}$. Deep (light) blue is the region that $ \sigma/m_\chi =0.1-1~\text{cm}^2/$g ($1-10~\text{cm}^2/$g) in dwarf galaxies, solving the aforementioned issues of CCDM. The red solid (dashed) line shows the curve where  $\sigma/m_\chi =0.1~\text{cm}^2/$g ($1~\text{cm}^2/$g) in the Milky Way. The phase space to the left of the red curve is excluded because the cross section is sufficiently large to smooth out the ellipticity of Milky Way to a degree inconsistent with observations. There is a bit of ambiguity 
regarding the value of the maximum $\sigma/m_\chi$ consistent with observations but it should be between $0.1-1~\text{cm}^2/$g~\cite{Tulin:2013teo}. 

\section{Stellar Hydrostatic Equilibrium}

An asymmetric dark star resembles in several aspects a neutron star. Both types of stars produce no energy by fusing nuclei in their cores. Therefore there is no radiation pressure present. The structure of the star is determined by the equilibrium between the Fermi pressure of the constituent elements and gravity.
In Newtonian dynamics the above condition takes the simple form
\begin{equation}
\frac{dP}{dr}=-\frac{GM \rho}{r^2}, \label{pres_newt}
\end{equation}
where $P$, $M$, and $\rho$ are the pressure, the mass and the density of the star at radius $r$. Additionally, the continuity of mass gives
\begin{equation}
\frac{dM}{dr}=4 \pi r^2 \rho. \label{mass_newt}
\end{equation}
The above two equations along with the equations of state (\ref{eqstate1}) and (\ref{eqstate2}) form a complete set of differential equations that can be solved numerically providing pressure, density and mass as a function of $r$. As mentioned earlier, in the absence of self-interactions, the equation of state takes a simple polytropic form $P=K \rho^{\Gamma}$ in both the relativistic and non-relativistic limits. This polytropic equation of state together with Eqs.~(\ref{pres_newt}) and (\ref{mass_newt}) reduce to the well known Lane-Emden equation with index $n=3/2,~3$ at the non-relativistic and relativistic limit respectively. However, as it has been seen in the case of neutron stars, the Newtonian approximation is not accurate enough and general relativity must be taken into account. We implement this by solving the Tolman-Oppenheimer-Volkoff equation together with  Eqs.~(\ref{mass_newt}), (\ref{eqstate1}) and (\ref{eqstate2}).

For completeness, let us briefly review how the  Tolman-Oppenheimer-Volkoff equation is obtained. We are seeking a solution of the Einstein field equation
$R_{\mu\nu} -  g_{\mu \nu} R/2 = 8\pi G T_{\mu \nu}$, in the presence of matter with an energy momentum tensor of an ideal liquid of the form
\begin{equation}
T_\mu^\nu  = \text{diag}\left[\rho,-P,-P,-P\right].
\end{equation}
If one plugs the following spherically symmetric metric into Einstein's field equations
\begin{equation}
ds^2=e^{\nu(r)}dt^2-e^{\lambda(r)}dr^2-r^2d\Omega^2, \label{eq:interiormetric}
\end{equation}
one finds the following set of equations that must be satisfied
\begin{align}
8\pi GP &= e^{-\lambda} \left(\frac{\nu'}{r}+\frac{1}{r^2} \right)-\frac{1}{r^2},\label{eq:derive nu}\\
8\pi G\rho &= e^{-\lambda} \left( \frac{\lambda'}{r}-\frac{1}{r^2}\right)+\frac{1}{r^2},\label{eq:derive mass}\\
\frac{dP}{dr}&=-\frac{(P+\rho) \nu'}{2}.\label{eq:derive pressure}
\end{align}
Requiring that the metric reduces to the empty space Schwarzschild solution at the boundary of the star, provides a solution for $\lambda (r)$ and $\nu (r)$ and Eq.~(\ref{eq:derive pressure}) takes the final form

\begin{equation}
\frac{dP}{dr} = -\frac{GM \rho}{r^2} \frac{\left[1+\frac{P}{\rho} \right] \left[1+ \frac{4\pi r^3 P}{M}\right]}{\left[1-\frac{2GM}{r}\right]}\label{eq:TOV}.
\end{equation}
This is the relativistic version of Eq.~(\ref{pres_newt}). We have solved the coupled system of Eqs.~(\ref{eq:TOV}), (\ref{mass_newt}), (\ref{eqstate1}) and (\ref{eqstate2}) and we have obtained the structure profile of asymmetric dark stars both for attractive and repulsive self-interactions. Relativistic effects can be quite significant and therefore it is compulsory to use the Tolman-Oppenheimer-Volkoff equation instead of the simpler Newtonian version.

\subsection{Hydrostatic Stability \label{HS}}
Solving the Tolman-Oppenheimer-Volkoff equation yields an equilibrium solution, which may be stable or unstable. We will briefly review the conditions that must be satisfied in order for a star to pass from stability to instability. We assume a constant chemical composition and constant entropy per DM particle.

\begin{figure}
\centering
\includegraphics[width=.4\textwidth]{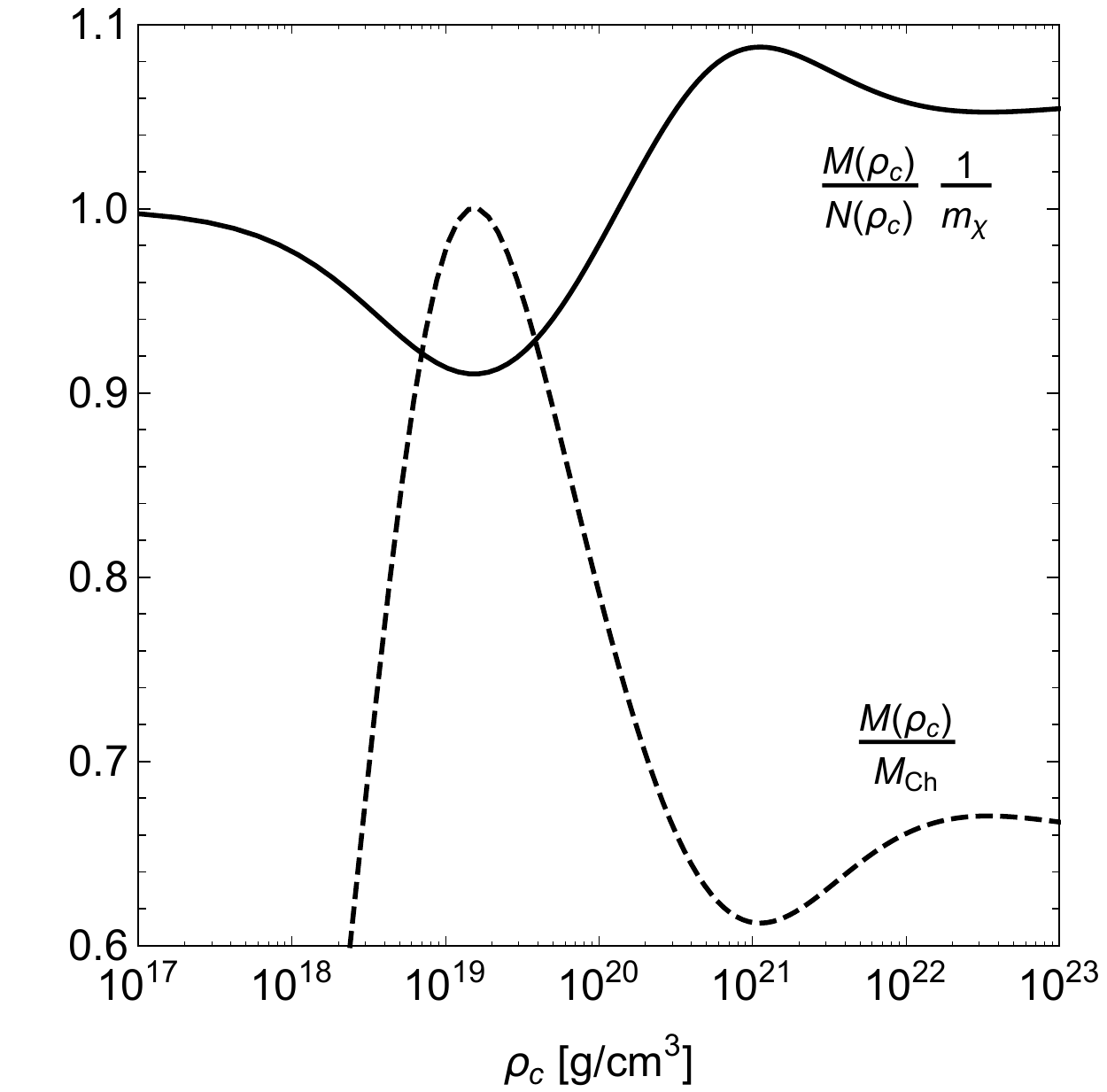}
\caption{$M/N m_\chi^{-1}$ and $M /M_\text{Ch}$ as a function of the central density. In this example the interactions are repulsive, $m_\chi =100$ GeV, $\mu=10$ MeV and $\alpha = 10^{-3}$. The central density at which the energy per particle is minimized coincides with the density at the Chandrasekhar mass in Fig. \ref{fig:MvsR} (a) (marked by a blue circle). For this reason the equilibrium configurations at higher densities are unstable. As expected, at low densities $M/N$ is simply $m_\chi$.}
\label{fig:MperN}
\end{figure}

The total mass-energy of the star is the integrated energy density
\begin{equation}
M(R)=\int_0^R 4\pi r^2 \rho dr,
\end{equation}
where $R$ is defined by $P(R)=0$. The number of constituent DM particles in the star is \cite{Weinberg:1972}
\begin{equation}
N(R)= \int_0^R 4\pi r^2 \left[1-\frac{2GM(r)}{r} \right]^{-1/2} n dr,
\end{equation}
where $n$ is the DM number density as a function of $r$.
A star of constant chemical composition and entropy per particle can only pass from stability to instability with respect to some particular radial normal mode, at a value of the central density $\rho_c$ for which we have \cite{Weinberg:1972}
\begin{equation}
\frac{\partial M}{\partial \rho_c} = 0, \quad \frac{\partial N}{\partial \rho_c} = 0.
\end{equation}
%For the star to be stable with respect to a change in the central density, the energy per particle $M/N$ must decrease, we expect that in the low density limit the mass $\lim_{\rho_c \to 0} M/N = m_\chi$ and for a %transition from stability to instability to occur we must have
These conditions are satisfied if we alternatively choose to satisfy simultaneously the first equation above and
\begin{equation}
\frac{\partial}{\partial \rho_\mathrm{c}} \left( \frac{M}{N}\right) = 0.
\end{equation}
Fig.~\ref{fig:MperN} shows $M/N$ (divided by $m_\chi^{-1}$) and $M$ (in units of the corresponding upper mass $M_\text{Ch}$) as a function of $\rho_c$ in the repulsive case with  $m_\chi =100$ GeV, $\mu=10$ MeV and $\alpha = 10^{-3}$. One can see that the points where $M/N$ minimizes and $M$ maximizes  coincide. This is a generic feature for all the star profiles we present in Figs.~ \ref{fig:MvsR} and \ref{fig:MvsR2}, i.e. the transition from stability to instability (collapse) takes place at the point where the mass maximizes.  

In the absence of Yukawa interactions, the critical relativity parameter at transition from stability to instability is $x_c \sim 0.8$ \cite{Shapiro:1983du}, which is independent of $m_\chi$. Yukawa interactions induce dependence of $x_c$ on $m_\chi$, $\mu$ and $\alpha$. 
%We will show that dark stars consisting of heavy DM can be non-relativistic, whereas light DM stars may be relativistic. In the repulsive Yukawa interacting scenario a large coupling increases the mass of the dark %star, yet decreases the central density; in this case one may have a peculiar situation where the equilibrium solutions exhibit large GR corrections while all particles follows a non-relativistic dispersion relation. On %the other hand if the interactions are attractive, a large coupling allows only for small configurations. From the equations of state \ref{eqstate1} and \ref{eqstate2} one can see that increasing $\alpha$ has the same %qualitative effect as decreasing $\mu$.

\section{Analytic Newtonian Approximation}
We would like to estimate the upper mass limit (Chandrasekhar mass) for asymmetric dark stars using first simple Newtonian arguments. The reader interested in the full relativistic results may skip this section and move to the next one. 
In order to understand the different regimes the dark star passes through before a collapse occurs we derive analytical solutions using a Newtonian approximation for gravity and assume a constant density. Comparing this approach to the full relativistic treatment of the Tolman-Oppenheimer-Volkoff equation, will provide an idea of how important general relativity effects are in different DM scenarios.

In this simplified picture we are going to minimize the energy of the system upon making some approximations. We use Newton's gravitational law (instead of general relativity) and we assume a uniform density of DM fermions with Yukawa interactions. The energy has three contributions, i.e. kinetic energy, gravitational potential energy and Yukawa potential energy
\begin{equation}
E=E_G + E_\text{kin} + E_\text{Y}.
\end{equation}
The self gravity contribution to the energy is
\begin{equation}
E_G = - \int_0^R \frac{G}{r} \frac{4}{3}\pi\rho r^3  \cdot 4\pi\rho r^2 dr = - \frac{3}{5}\frac{Gm_\chi^2N^2}{R},
\end{equation}
where $N$ is the total number of particles and $R$ the radius of the star.
The kinetic energy is found by multiplying Eq.~(\ref{eq:kinetic density}) by the volume
\begin{equation}
E_\text{kin} =  \frac{2\pi g_s}{3} R^3 m^4 \xi\left( \frac{p_\text{F}}{m_\chi}\right),
\end{equation} 
where $p_F$ is the Fermi momentum of the particles.
In the relativistic and non-relativistic limits, the kinetic energy assumes simple polynomial forms in $1/R$:
\begin{equation}
E_\text{kin}= 
\begin{dcases}
    m_\chi N+\frac{g_s}{15\pi} \left( \frac{9\pi}{2g_s}\right)^{5/3}\frac{N^{5/3}}{m_\chi R^2}& \text{if } \; \; p_\text{F}\ll m_\chi,\\
    \frac{g_s}{6\pi} \left( \frac{9\pi}{2g_s}\right)^{4/3} \frac{N^{4/3}}{R}              & \text{if } \; \; p_\text{F}\gg m_\chi.
\end{dcases}
\end{equation}
In our approximation we define a radius $R_\text{rel}$ such that $p_\text{F}(R_\text{rel}) = m_\chi$, 
\begin{equation}
R_\text{rel} = \left( \frac{9\pi}{2g_s}\right)^{1/3}\frac{N^{1/3}}{m_\chi}.
\end{equation}
When $R \ll R_\text{rel}$ the system is relativistic, while in the opposite limit ($R \gg R_\text{rel}$) the system can be treated as non-relativistic.

The Yukawa energy for the entire system of $N$ particles is more complicated to derive than the other terms in the energy. Starting from the potential between two particles in Eq. (\ref{eq:Yukwabetween2particles}), one can find the potential by integrating the contributions of the shells of a homogeneous sphere of charges ~\cite{Yukawaderivation}. The final interaction energy reads 
\begin{equation}
E_\text{Y} = \pm \frac{3}{4} \frac{\alpha N^2}{\mu^5R^6} \left[2 \mu^3R^3 -3\mu^2R^2 + 3 - 3(1+\mu R)^2 e^{-2\mu R} \right].
\end{equation}
This Yukawa potential energy also assumes simple polynomial forms in $1/R$ in the long and short range limits where $\mu R\ll 1$ and $\mu R \gg 1$ respectively
\begin{equation}
E_\text{Y}= 
\begin{dcases}
   \pm \frac{3}{5}\frac{\alpha N^2}{ R}& \text{if } \; \; \mu R\ll 1,\\
   \pm \frac{3}{2}\frac{\alpha N^2}{\mu^2 R^3}               & \text{if } \; \; \mu R\gg 1.
\end{dcases}
\end{equation}
It is no surprise that the energy in the long range regime is independent of the mediator mass and proportional to $1/R$ since it resembles the Coulomb potential. In the short range limit where $\mu R \gg 1$, the exponential suppression of the potential is counterbalanced by the number of close neighbours, thus leading to a $R^{-3}$ overall dependence. 
Given the above, we identify four distinct regimes that can be realised by the system: 
\begin{enumerate}
\item {Non-relativistic, short range: $p_\text{F} \ll m_\chi$ and $\mu R \gg 1$:
\begin{align}
E(N,R) = - & \frac{3}{5}\frac{Gm_\chi^2N^2}{R}+ m_\chi N+ \notag\\
&\frac{g_s}{15\pi} \left( \frac{9\pi}{2g_s}\right)^{5/3}\frac{N^{5/3}}{m_\chi R^2}  \pm \frac{3}{2}\frac{\alpha N^2}{\mu^2 R^3}. \label{eq:case1}
\end{align}
}
\item {Non-relativistic, long range: $p_\text{F} \ll m_\chi $ and $\mu R \ll 1$:
\begin{align}
E(N,R) =  -& \frac{3}{5}\frac{Gm_\chi^2N^2}{R}+m_\chi N+ \notag\\
&\frac{g_s}{15\pi} \left( \frac{9\pi}{2g_s}\right)^{5/3}\frac{N^{5/3}}{m_\chi R^2}  \pm \frac{3}{5}\frac{\alpha N^2}{ R}. \label{regime2}
\end{align}
}
\item {Relativistic, short range: $p_\text{F} \gg m_\chi $ and $\mu R \gg 1$:
\begin{equation}
E(N,R) =  - \frac{3}{5}\frac{Gm_\chi^2N^2}{R}+ \frac{g_s}{6\pi} \left( \frac{9\pi}{2g_s}\right)^{4/3} \frac{N^{4/3}}{R} \pm \frac{3}{2}\frac{\alpha N^2}{\mu^2 R^3}.\label{regime3}
\end{equation}}

\item {Relativistic, long range: $p_\text{F} \gg m_\chi $ and $\mu R \ll 1$:
\begin{equation}
E(N,R) =- \frac{3}{5}\frac{Gm_\chi^2N^2}{R}+   \frac{g_s}{6\pi} \left( \frac{9\pi}{2g_s}\right)^{4/3} \frac{N^{4/3}}{R}\pm \frac{3}{5}\frac{\alpha N^2}{ R}.\label{regime4}
\end{equation}}
\end{enumerate}

We would like to estimate the condition for gravitational collapse of the objects, thus determining an upper mass limit for the asymmetric fermionic dark stars, similar to the Chandrasekhar limit in the case of white dwarfs. We assume that the star starts always from regime 1. However, depending on the parameters, the star might pass from different regimes before the collapse  takes place. In practice we found two possibilities that take place most of the time. In the case of attractive interactions the star can move from regime 1 to  3 and collapse or from regime 1 to 3 to 4 and then collapse. For the repulsive potential the system goes directly from regime 1 to collapse. In the next subsections we will analyse each regime in greater detail.

\subsection*{Regime 1: Non-relativistic, short range}
 We rewrite Eq.~(\ref{eq:case1}) in terms of  new constants $A,B,C$ keeping the explicit dependence of $N$ and $R$.
\begin{equation*}
E(N,R) = - A \frac{N^2}{R}+B\frac{N^{5/3}}{R^2} + C\frac{ N^2}{ R^3}.
\end{equation*}
The derivative $dE/dR=0$ gives two extrema
\begin{equation}
R_\pm = \frac{B}{A} N^{-1/3} \pm \sqrt{\left( \frac{B}{A}\right)^2 N^{-2/3} + \frac{3C}{A}}.
\end{equation}
For an attractive potential ($C<0$) $R_+$ is a minimum and $R_-$ a maximum. If one increases the value of $N$,  there is a particular value where $R_+=R_-$ and the system collapses because there is no stable solution. This value of $N$ is
\begin{equation}
N_\text{max} = \left(-\frac{B^2}{3CA} \right)^{3/2} = \frac{9\pi^2}{40g_s^2} \sqrt{\frac{3}{10}} \frac{1}{\alpha^{3/2}} \left( \frac{\mu}{m_\chi }\right)^3\left( \frac{M_\text{P}}{m_\chi }\right)^3.\label{eq:nmax}
\end{equation}
When $N=N_{\text{max}}$, one can find that $\mu R_+(N_{\text{max}})=\sqrt{15/2}\sqrt{\alpha} M_\text{P}/m_\chi  \gg 1$ for the whole parameter space we examine here. Therefore when the system has accumulated $N_{\text{max}}$ particles, Yukawa forces are still short range. Since the energy has no minimum, $R$ keeps dropping until $R_{\text{rel}}$ or the Schwarzschild radius $R_\text{s}=2GNm_\chi $ is reached depending on which one is the larger. If $R_{\text{rel}}>R_\text{s}$, one can estimate $\mu R_{\text{rel}}(N_{\text{max}}) \simeq (2.6 /g_s)(1/ \sqrt{\alpha})M_\text{P} \mu^2/m_\chi^3\gg1$. This means that for the parameter space considered here the system reaches $R_{\text{rel}}$ before the Yukawa force becomes long range. Therefore the system will go to regime 3 or 4 and not 2. If $R_\text{s}>R_{\text{rel}}$, the system collapses to a black hole before the particles become relativistic. In this case $\mu R_\text{s}\gg1$ for the parameter space we consider, and therefore the system forms a black hole without reaching the long range limit of the Yukawa force. 

In the case where the Yukawa potential is repulsive ($C>0$), $R_-< 0$ (so it is unphysical) and $R_+$ is a global minimum. The upper mass limit for this star can be found by setting  $R_+(N) = R_\text{s}(N)$. There is always an appropriate $N$ that satisfies this because  $R_+(N)$ decreases with $N$ while $R_\text{s}(N)$ increases. For the parameter space we consider here, we found that the Schwarzschild radius is encountered while still in the non-relativistic regime. The smallest possible value of $R_+$ (which occurs when $N\rightarrow \infty$) is $R_\text{s} = \sqrt{3C/A}$. One can see that for the parameter space considered here, this line is crossed first by $R_\text{s}$ and then by $R_{\text{rel}}$. It can be easily seen that this is also true even for a finite $N$, thus the star collapses to a black hole before the constituents become relativistic. In addition, one can show that the aforementioned asymptotic value satisfies  $\mu R_\text{s} \gg 1$ and therefore  the interactions are not long range.

\subsection*{Regime 2: Non-relativistic, long range}
We rewrite Eq.~(\ref{regime2}) with new constants $A,B,C$ as 
\begin{equation*}
E(N,R) =  - A\frac{N^2}{R}+B\frac{N^{5/3}}{R^2}  +C\frac{ N^2}{ R}.
\end{equation*}
The relevant minimum is found at
\begin{equation}
R_\text{min} = \frac{2B}{A-C}N^{-1/3} = \left( \frac{9\pi}{2g_s}\right)^{2/3}\frac{1}{m_\chi  N^{1/3}} \left[\left(\frac{m_\chi }{M_\text{P}}\right)^2 \mp \alpha\right]^{-1}.
\end{equation}
If the Yukawa potential is attractive ($C<0$), there is always a stable minimum. On the other hand in the case of repulsive potential $C>0$, the strength of the coupling is bound by $\alpha< m_\chi ^2/M_\text{P}^2$, if a stable minimum is to exist at finite positive $R$. In practice for the considered parameter space, we found that no dark star passes through this regime.

\subsection*{Regime 3: Relativistic, short range}
Once again we rewrite Eq.~(\ref{regime3}) in terms of new constants $A,B,C$ as
\begin{equation}
E(N,R) =  - A\frac{N^2}{R}+ B \frac{N^{4/3}}{R} +C\frac{ N^2}{ R^3}
\end{equation}
and find the extremum at
\begin{equation}
R_\text{ext} = \sqrt{\frac{3C}{A-BN^{-2/3}}}.
\end{equation}
In the case of an attractive potential ($C<0$), this extremum is a maximum as long as $N$ is smaller than
\begin{equation}
N_\text{ext} = \left( \frac{B}{A}\right)^{3/2} = \frac{15 }{8} \sqrt{\frac{5\pi}{2g_s}}\left(\frac{M_\text{P}}{m_\chi }\right)^3.
\end{equation}
For $N> N_\text{ext}$ the energy is monotonically increasing as a function of $R$, and therefore the star collapses. For $N< N_\text{ext}$ the energy  increases between 0 and $R_\text{ext}$ and decreases from $R_\text{ext}$ to $\infty$. One can compare  $N_\text{max}$ from Eq.~(\ref{eq:nmax}) and $N_{\text{ext}}$
\begin{equation}
\frac{N_\text{max}}{N_\text{ext}} = \frac{3 \sqrt{3} \pi^{3/2}}{125\alpha^{3/2}}\frac{1}{g_s^{3/2}} \left( \frac{\mu}{m_\chi }\right)^3.
\end{equation}
If the star passes from regime 1 to regime 3, it collapses if $N_\text{max}>N_\text{ext}$. If on the other hand $N_\text{max}<N_\text{ext}$, the collapse proceeds as long as  $R_{\text{rel}}(N_{\text{max}})< R_\text{ext}(N_{\text{max}})$.

In the case of repulsive Yukawa potential ($C>0$), the potential is monotonically decreasing for $N<N_\text{ext}$. This means that the star remains in  regime 1. Once  $N> N_\text{ext}$ there is a stable minimum at $R_\text{ext}$. However, as we pointed out in the discussion in "Regime 1", the Schwarzschild radius is met before this.

\subsection*{Regime 4: Relativistic, long range}
In this case every term scales as $1/R$. Rewriting Eq.~(\ref{regime4}) in terms of new $A,B,C$ constants we get
\begin{equation}
E(N,R) =-A\frac{N^2}{R}+  B \frac{N^{4/3}}{R}+C\frac{ N^2}{ R}.
\end{equation}
The critical number of particles is
\begin{equation}
N_\text{crit} = \left( \frac{B}{A-C}\right)^{3/2} = \frac{15}{8}\sqrt{\frac{5\pi}{2g_s}} \left[\left( \frac{m_\chi }{M_\text{P}} \right)^2 \mp \alpha\right]^{-3/2}.
\end{equation}
For $N>N_\text{crit}$ the star collapses. In the case of repulsive interactions with $\alpha>m_\chi ^2/M_\text{P}^2$, no collapse can take place for any value of $N$. As we mentioned in ``Regime 1", in case of repulsive interactions we never enter this regime because the particle never becomes relativistic.

\section{Results}\label{sec:results}
In this section we present the full relativistic results after solving numerically the system of Eqs.~(\ref{eq:TOV}), (\ref{mass_newt}), (\ref{eqstate1}) and (\ref{eqstate2}). 
The algorithm we use to solve the relativistic hydrostatic equilibrium is the following:
\begin{enumerate}
\item Set initial conditions $M_0=M(r=0)=0$ and $P_0=P(r=0) = P(x_0)$, with $x_0=x_\text{c}$ being the relativity measure in the center of the star defined below Eq.~(\ref{7}).

\item Integrate one step of Eq. (\ref{mass_newt}) to get $M_1=M_0+\int dM$ using the equation of state for $\rho_0 = \rho(x_0)$. Then integrate one step of Eq. (\ref{eq:TOV}) to obtain $P_1 = P_0 + \int dP$. For this value of $P_1$, one can obtain the corresponding $x_1$ from the equation of state (\ref{eqstate1}). 

\item Repeat $i$ times the above step to obtain $M_{i}$ and $x_i$.

\item Identify the $R$ where $P(R)=0$. This defines the radius of the star. Correspondingly the mass of the star is $M(R)$.
\end{enumerate}
For each set of DM parameters we find the Chandrasekhar mass by scanning over $x_\text{c}$ and identifying the largest total mass.

\subsection{Mass-Radius relations}
\begin{figure*}
\begin{tabular}{cc}

  \includegraphics[width=.35\textwidth]{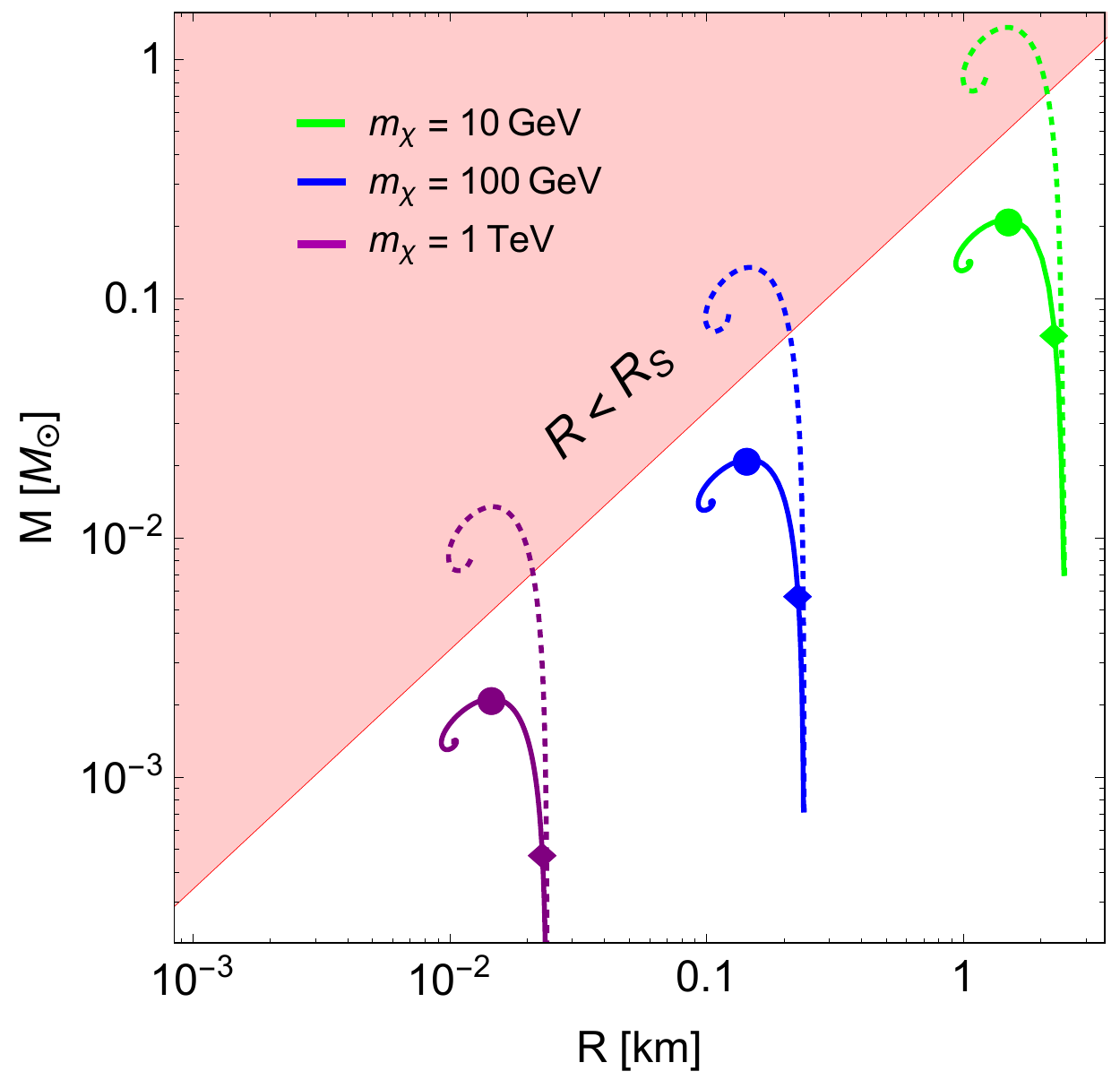} &   \includegraphics[width=.35\textwidth]{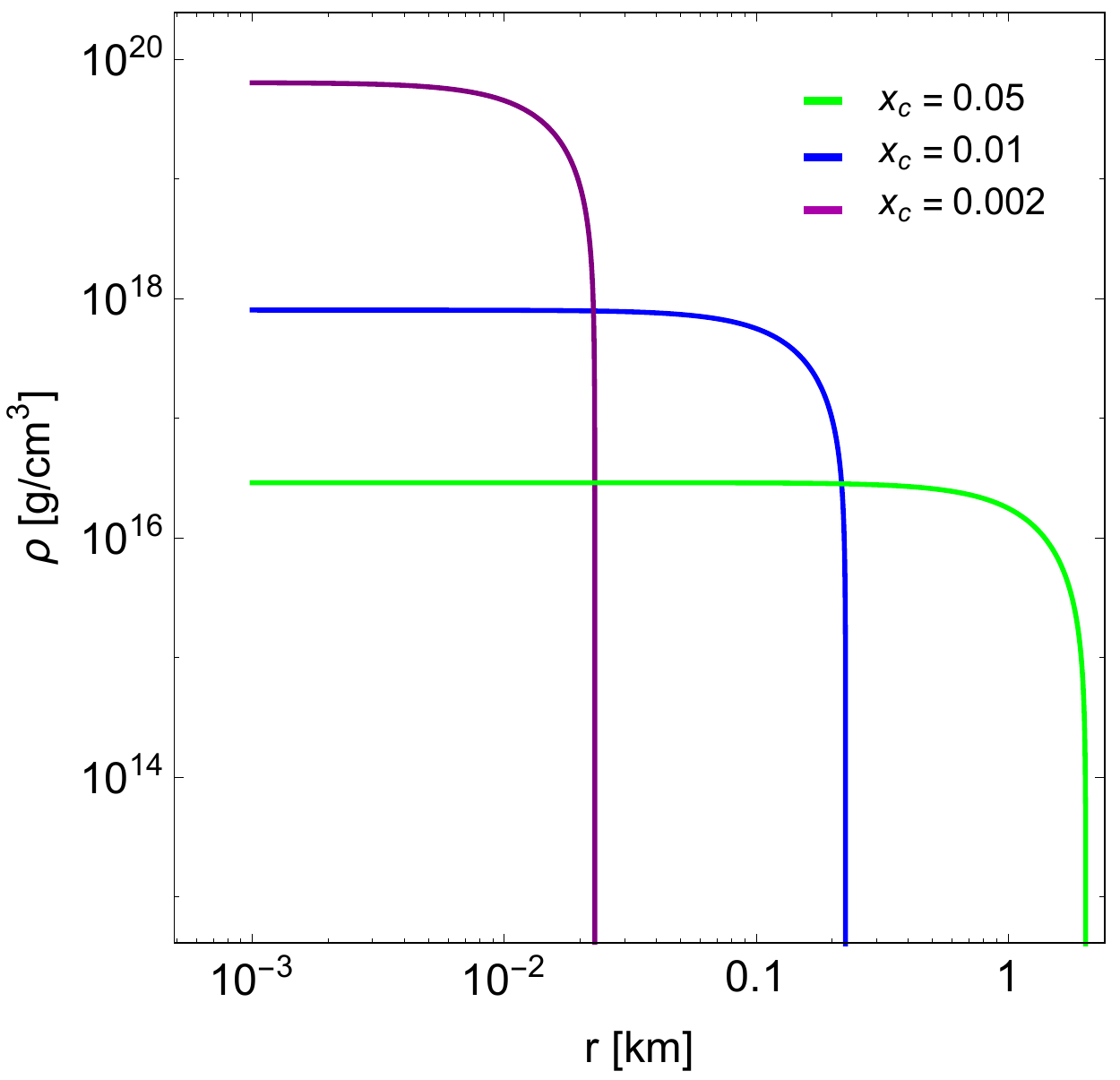} \\
(a) $M(R)$ for repulsive interactions & (b) $\rho(r)$ for repulsive interactions\\[6pt]

    \includegraphics[width=.35\textwidth]{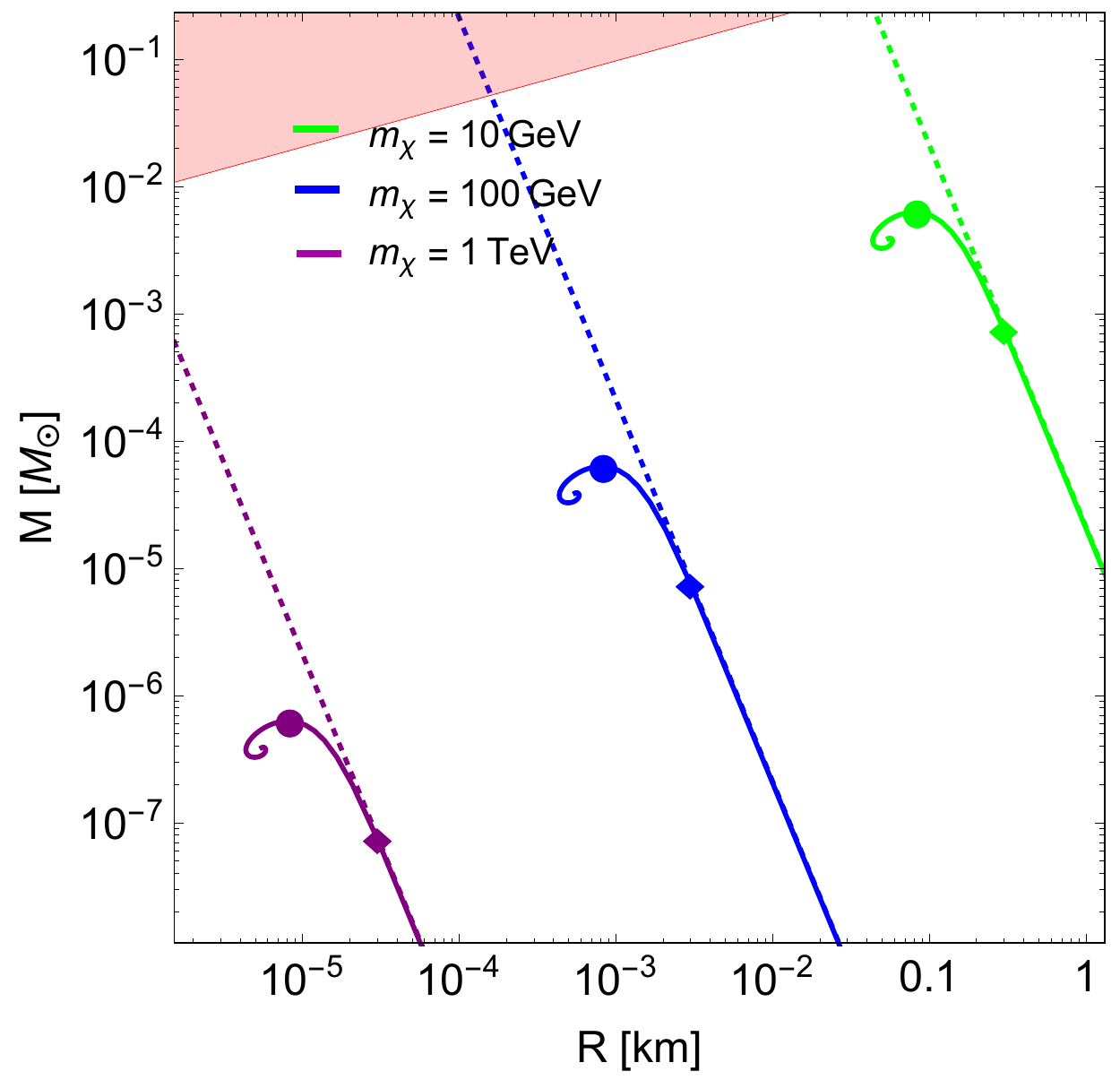} &   \includegraphics[width=.35\textwidth]{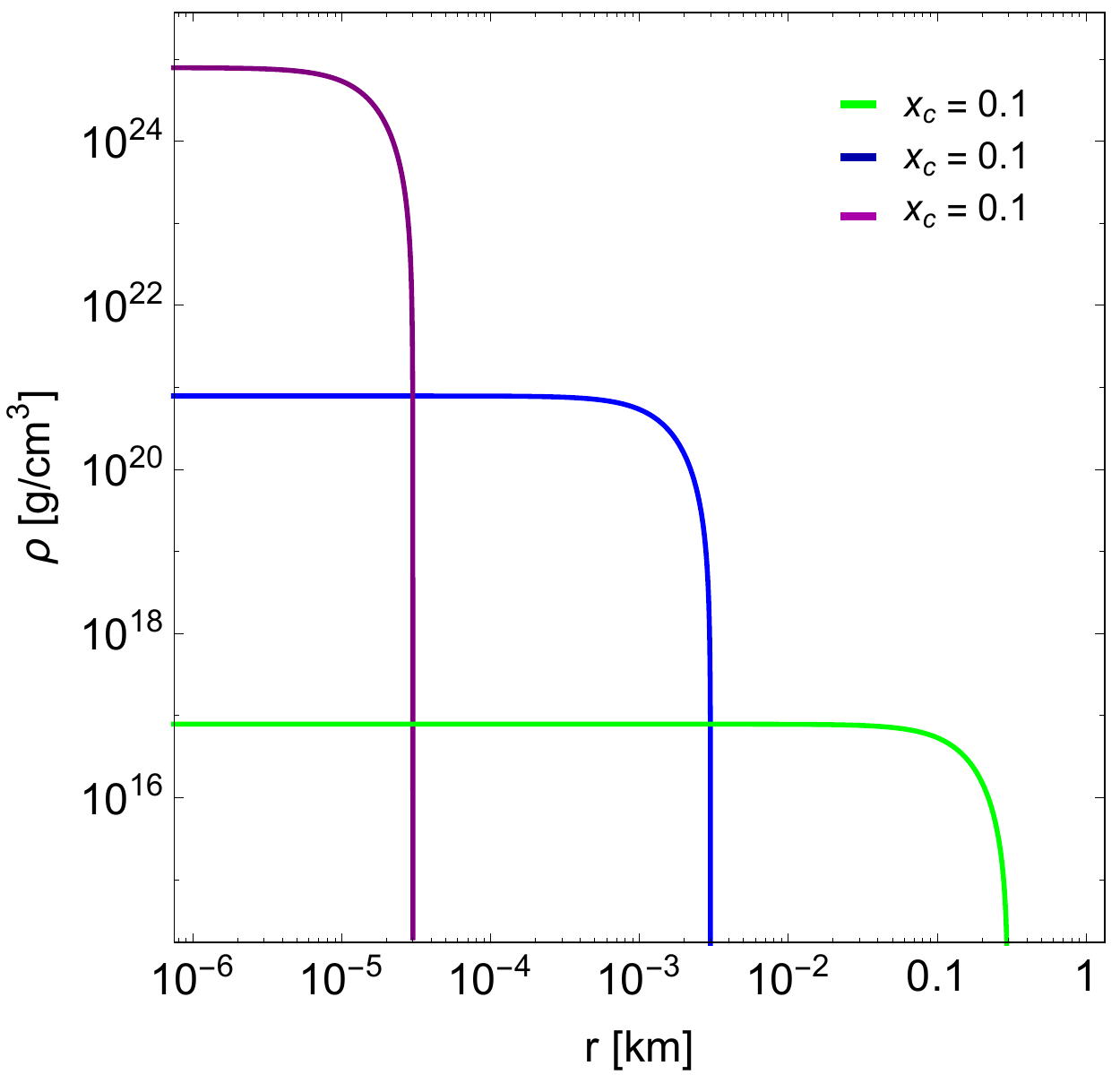} \\
 (c) $M(R)$ in the absence of interactions & (d) $\rho(r)$ in the absence of interactions \\[6pt]

  \includegraphics[width=.35\textwidth]{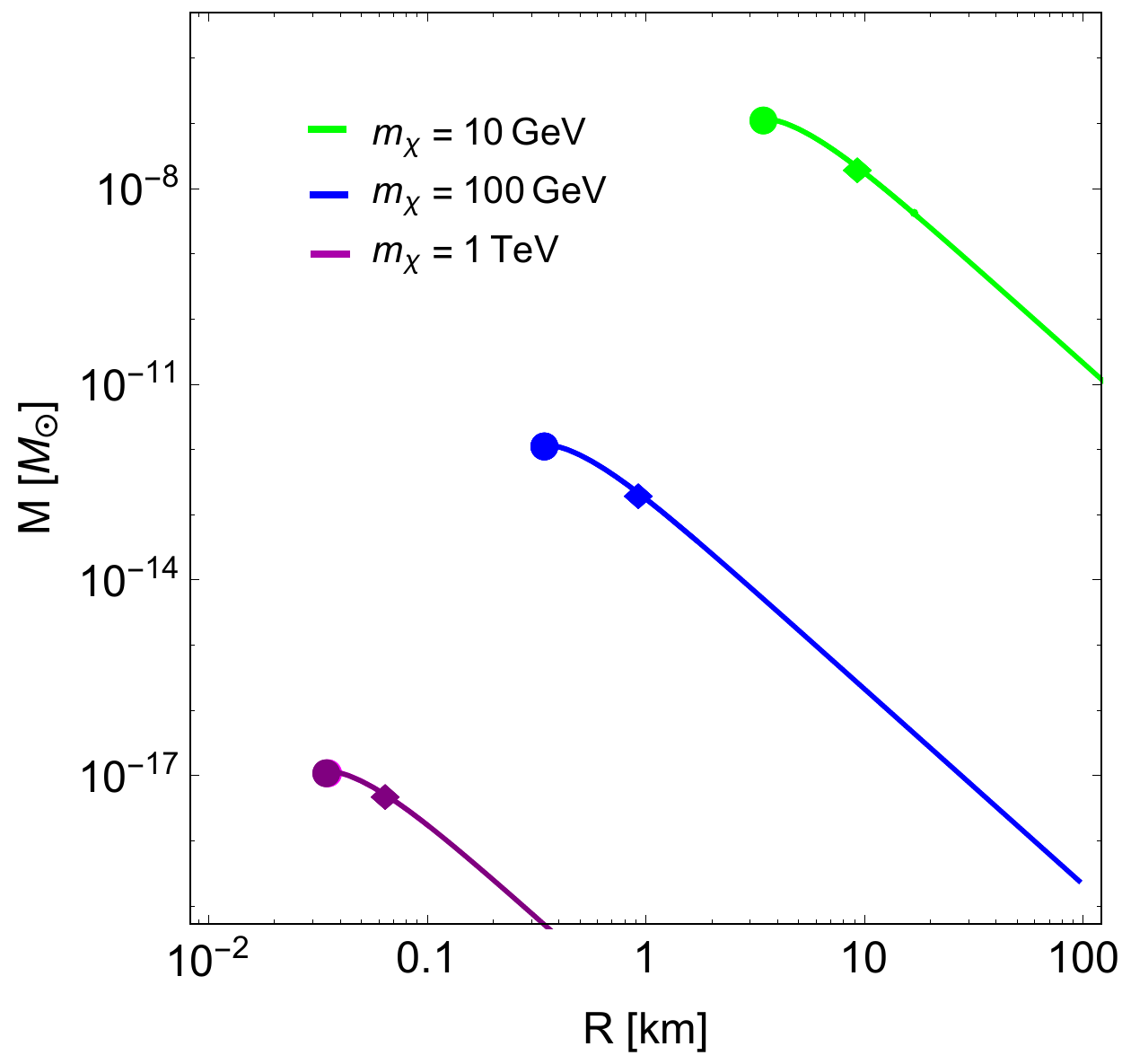} &   \includegraphics[width=.35\textwidth]{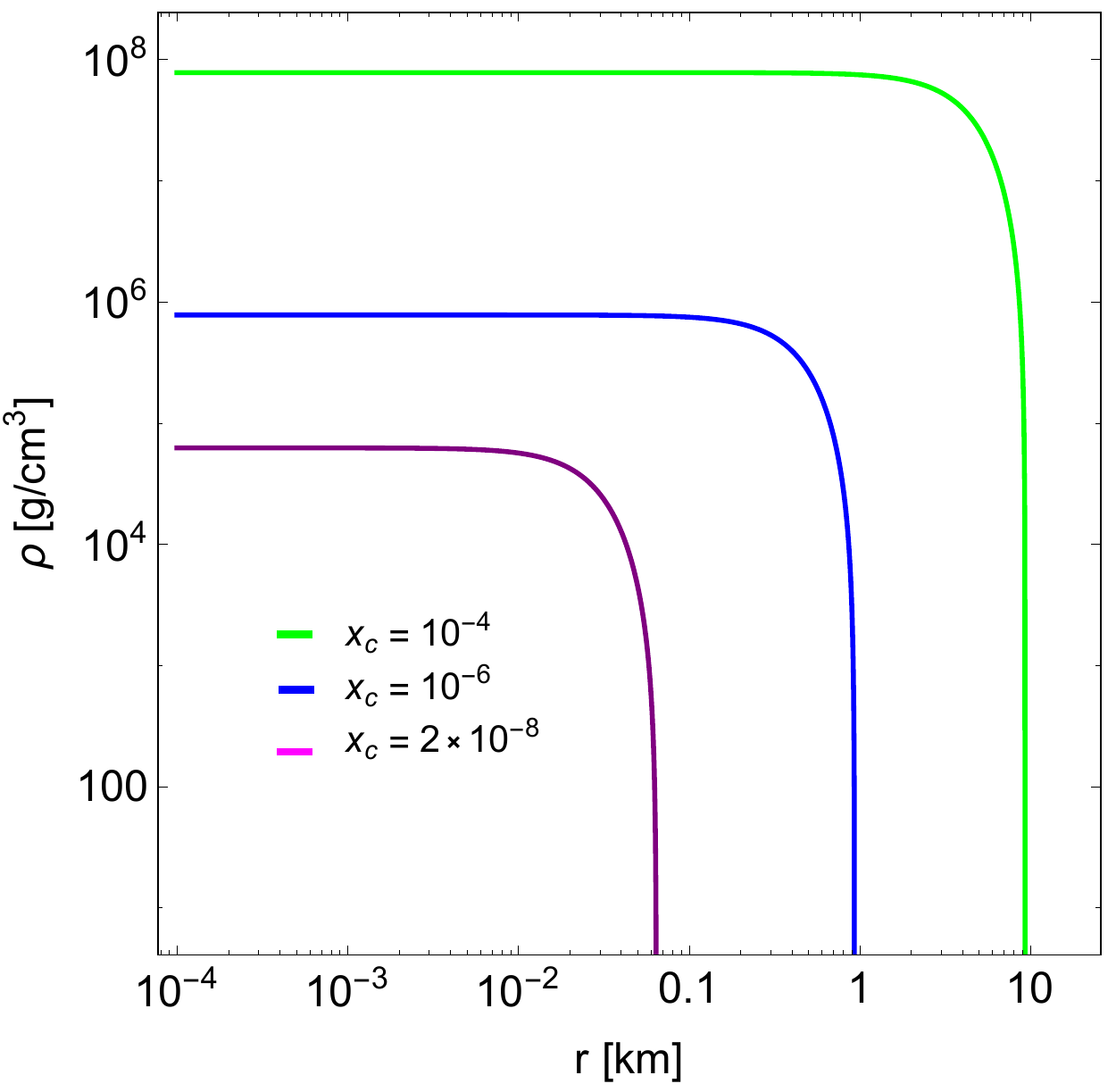} \\
 (e) $M(R)$ for attractive interactions & (f) $\rho(r)$ for attractive interactions \\[6pt]
  
\end{tabular}
\caption{In the left panels we show dark star mass vs radius relations with DM mass $m_{\chi}=$ 10 GeV (Green), 100 GeV (blue), 1 TeV (purple). Upper, middle and bottom panels correspond to repulsive, no-interactions and attractive interactions respectively. We have fixed $ \mu= 10$ MeV and $\alpha = 10^{-3}$. Solid curves represent full relativistic solutions while dashed curves represent Newtonian gravity ones. The circles represent the Chandrasekhar masses and the diamonds represent stars with their density profiles  plotted as a function of the radius in the corresponding right panels. In the red regions $R<R_{\text{s}}$. In the attractive interaction scenario, the Newtonian solutions lie on top of the relativistic ones.}
\label{fig:MvsR}
\end{figure*}

We will present now the mass vs radius relations of the stable dark star configurations.
In the left panels of Fig. \ref{fig:MvsR} we show mass vs radius relations for the three generic cases: repulsive interactions (upper panel), no DM self-interactions (middle panel) and attractive interactions (bottom panel). We have chosen a coupling $\alpha=10^{-3}$, mediator mass $\mu=10$ MeV and three different cases of DM mass of 10, 100 and 1 TeV.
 For each $M(R)$ profile, we mark the upper stable mass (Chandrasekhar mass) by a circle. Note that star configurations with radii larger than the one that corresponds to the Chandrasekhar mass are stable, while configurations with smaller radii are unstable. We refer the reader to the discussion of stability in subsection \ref{HS} and to 
the example of Fig. \ref{fig:MperN} which shows that the Chandrasekhar mass is indeed the last stable configuration as density increases. Note also that the dashed lines represent dark star configurations where the non-relativistic version of the Tolman-Oppenheimer-Volkoff equation is used. As it can be seen, general relativity effects can be quite significant.

 For each mass vs radius curve that we show in the left panels of Fig. \ref{fig:MvsR}, apart from the Chandrasekhar mass, we mark in addition another point by a diamond.  In the respective right panels, we show the density profile of the dark star that corresponds to the diamond. In the scenarios of repulsive and no self-interactions, the equilibrium solutions feature a spiral structure in the unstable region of the $M(R)$ curves. This feature is absent in the attractive scenario since the pressure becomes negative for some $x$, and this $x$ is encountered before the spiral appears. One can notice  (as it is expected) that heavier DM particles form lighter and more compact dark stars.

\begin{figure*}
\begin{tabular}{cc}

  \includegraphics[width=.35\textwidth]{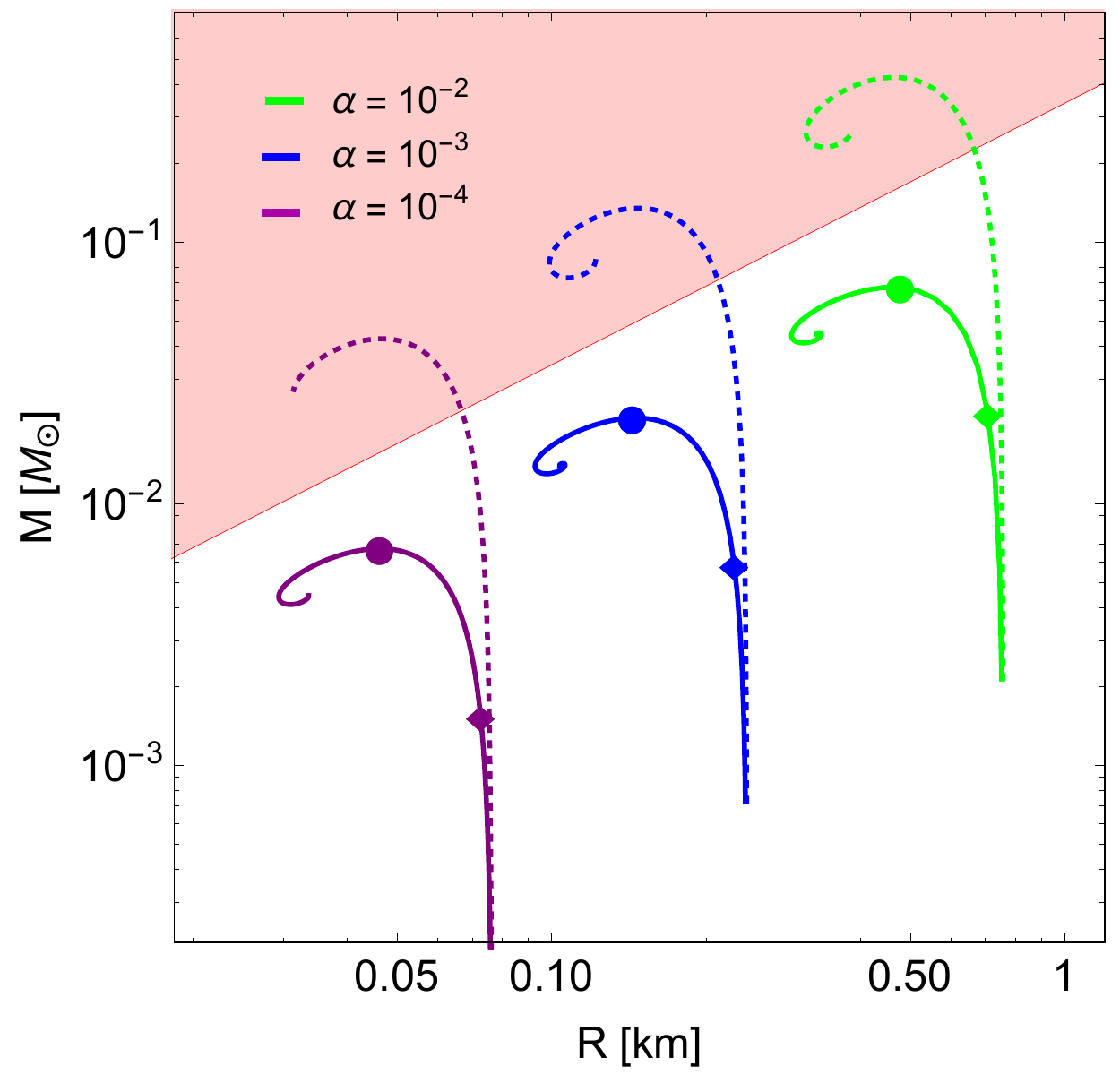} &   \includegraphics[width=.35\textwidth]{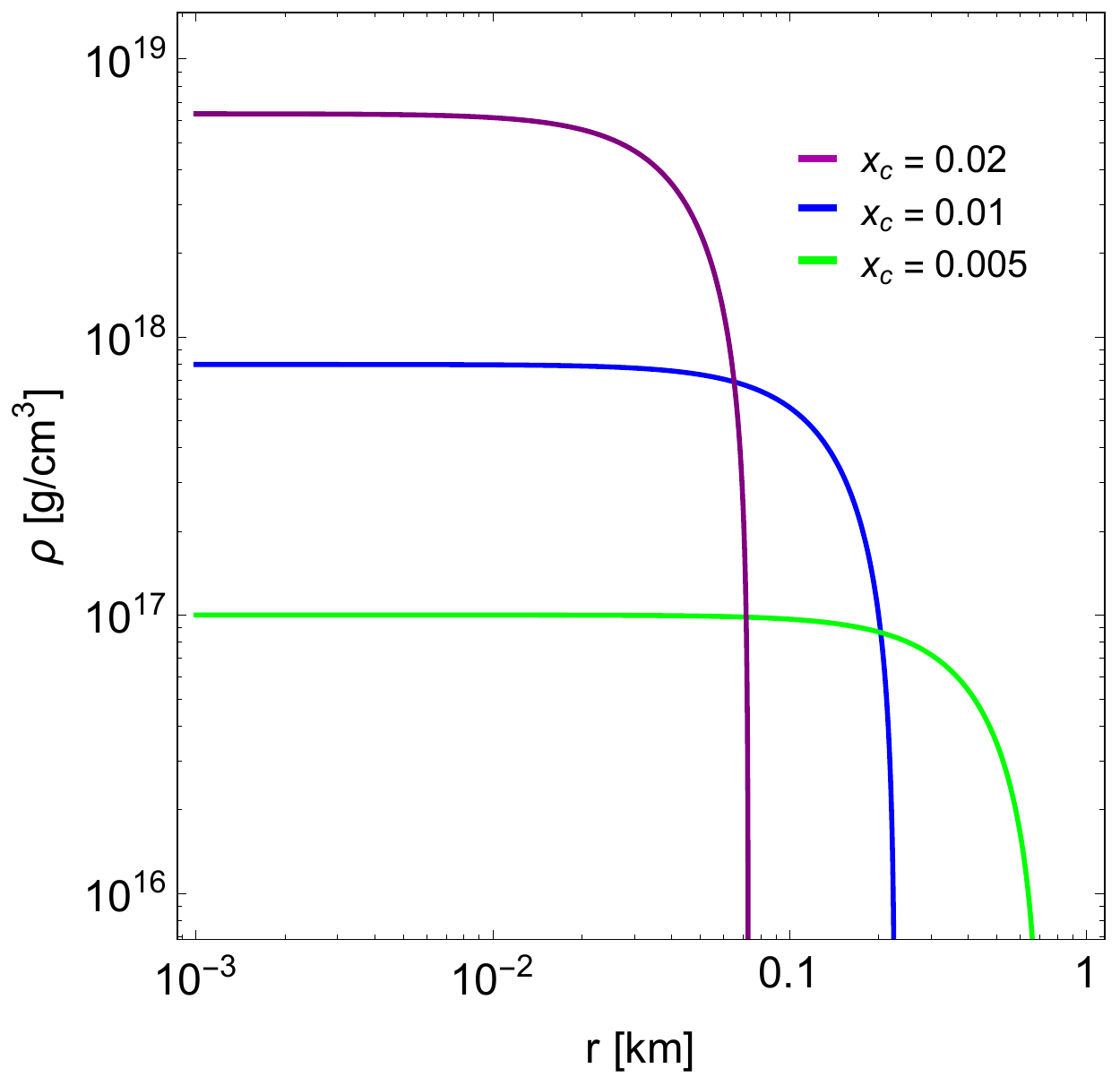} \\
(a) $M(R)$ for repulsive interactions & (b) $\rho(r)$ for repulsive interactions\\[6pt]

2  \includegraphics[width=.35\textwidth]{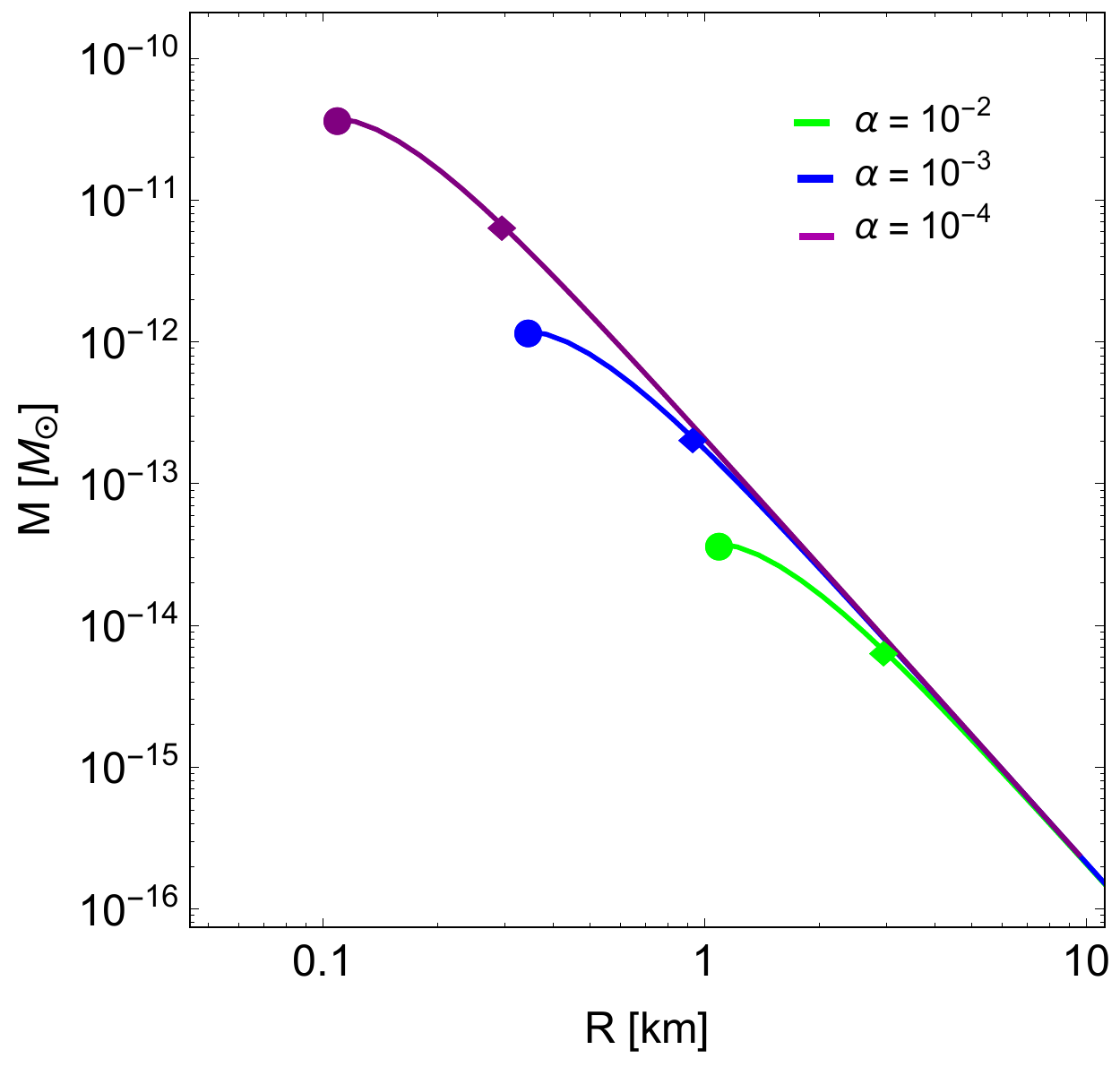} &   \includegraphics[width=.35\textwidth]{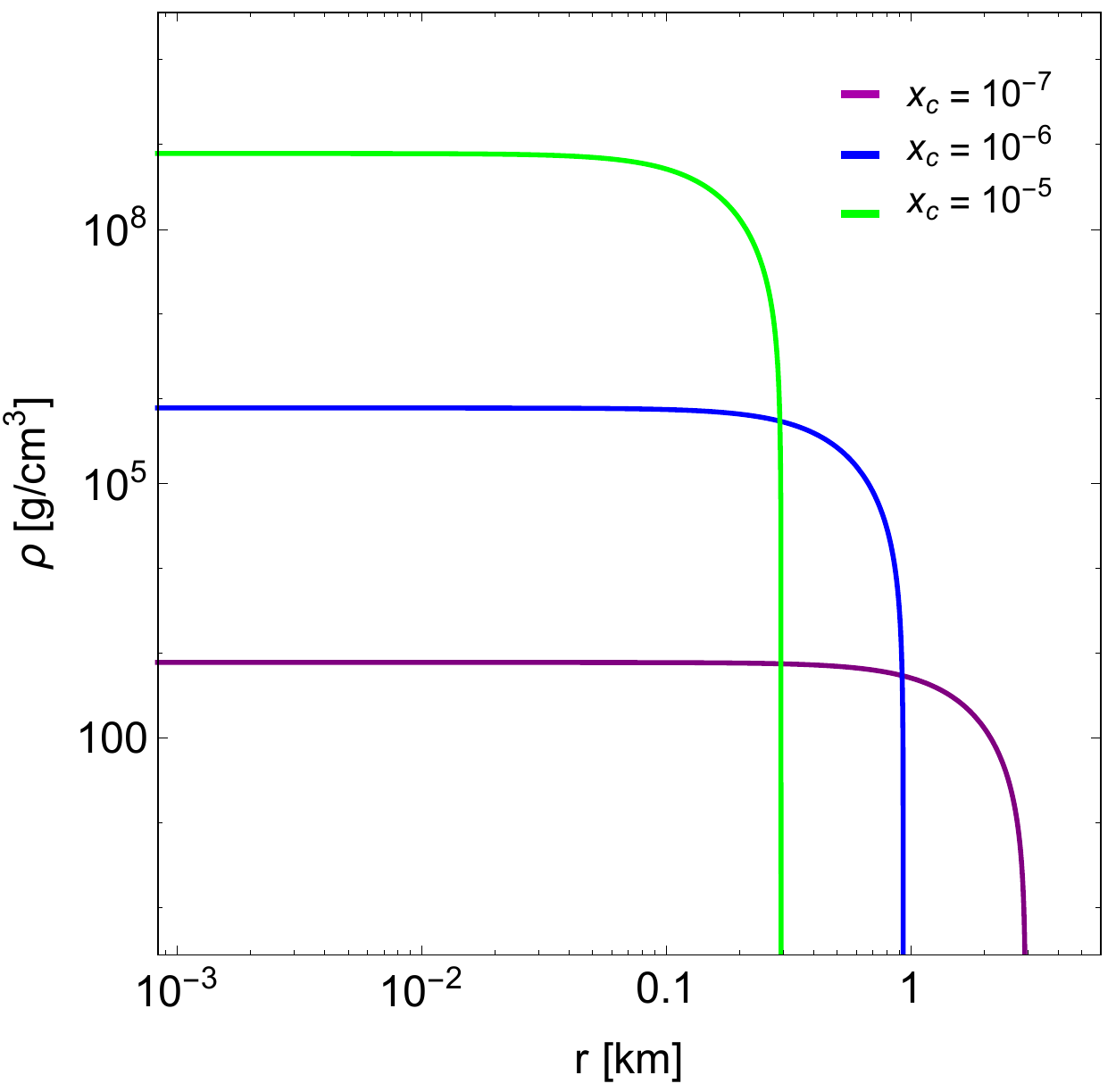} \\
 (c) $M(R)$ for attractive interactions & (d) $\rho(r)$ for attractive interactions \\[6pt]
  
\end{tabular}
\caption{In the left panels we show dark star mass vs radius relations for repulsive (upper panel) and attractive (right panel) for three different values of $\alpha= 10^{-2}$ (green), $\alpha = 10^{-3}$ (blue), and  $\alpha = 10^{-4}$ (purple). We have set $m_\chi = 100$ GeV, and $\mu = 10$ MeV. For the star configurations of the diamond points, we show the corresponding density profiles on the respective right panels. }
\label{fig:MvsR2}
\end{figure*}

 In Fig. \ref{fig:MvsR2} we show again mass vs radius relations and corresponding density profiles for the diamond points (as in Fig. \ref{fig:MvsR}) for three values of $\alpha=10^{-2},$ $10^{-3}$ and $10^{-4}$, for fixed $m_{\chi}=100$ GeV and $\mu=10$ MeV for repulsive and attractive interactions. Note from Eqs.~(\ref{eqstate1}) and (\ref{eqstate2}) that the Yukawa contribution to the density and the pressure of the star is unchanged  under the scaling $\mu \to q \mu$ and $\alpha \to q^2\alpha$. Therefore it is sufficient to fix either $\alpha$ and $\mu$ and scan the phase space of the other one. One can notice that increasing the coupling of repulsive Yukawa interactions leads to larger stars, although DM particles in these stars are non-relativistic  ($x_c\ll 1$). Despite that, the general relativity effects  are large.

\subsection{Chandrasekhar Mass}
\begin{figure*}
\begin{tabular}{cc}
  \includegraphics[width=.35\textwidth]{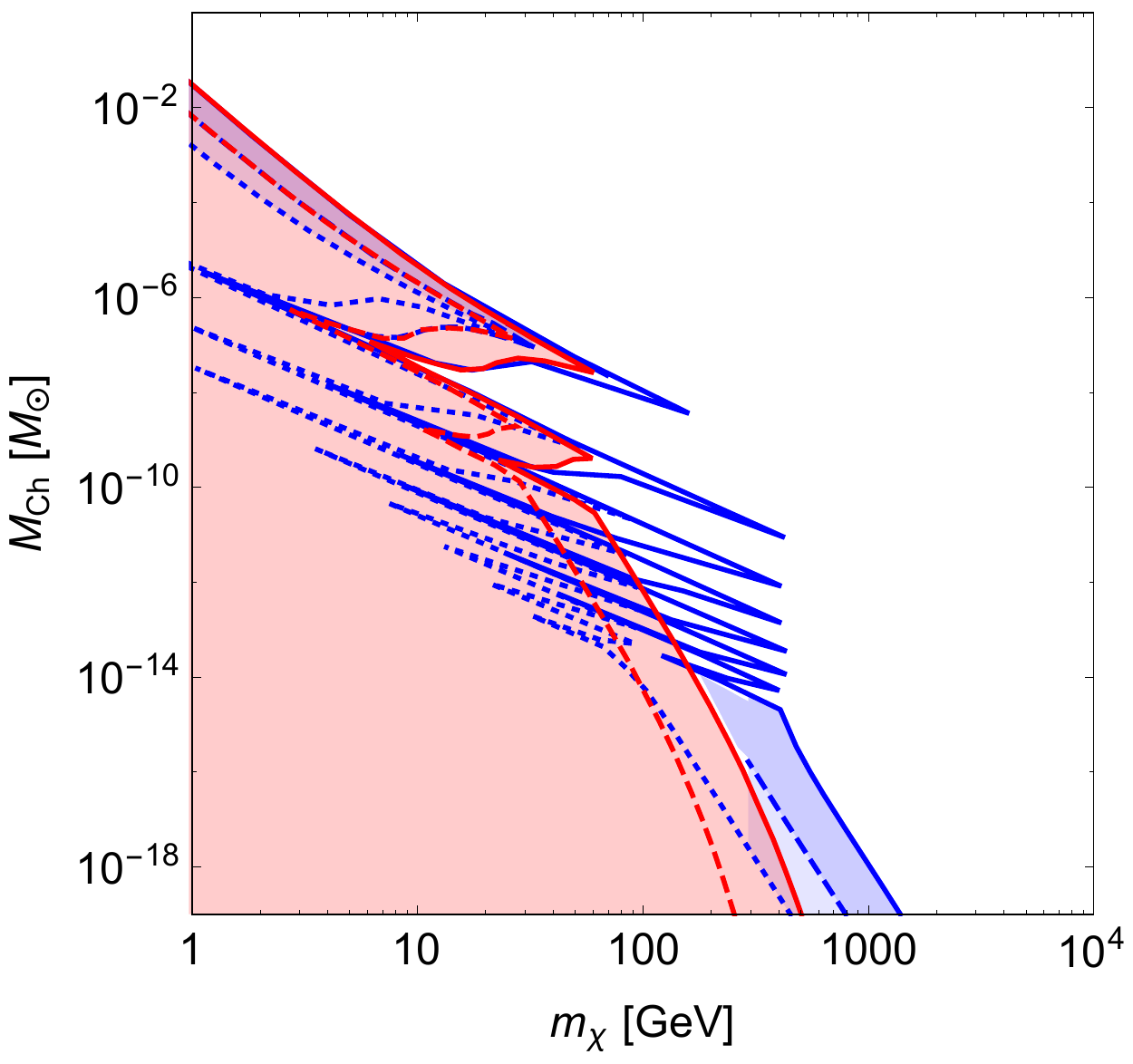} &   \includegraphics[width=.35\textwidth]{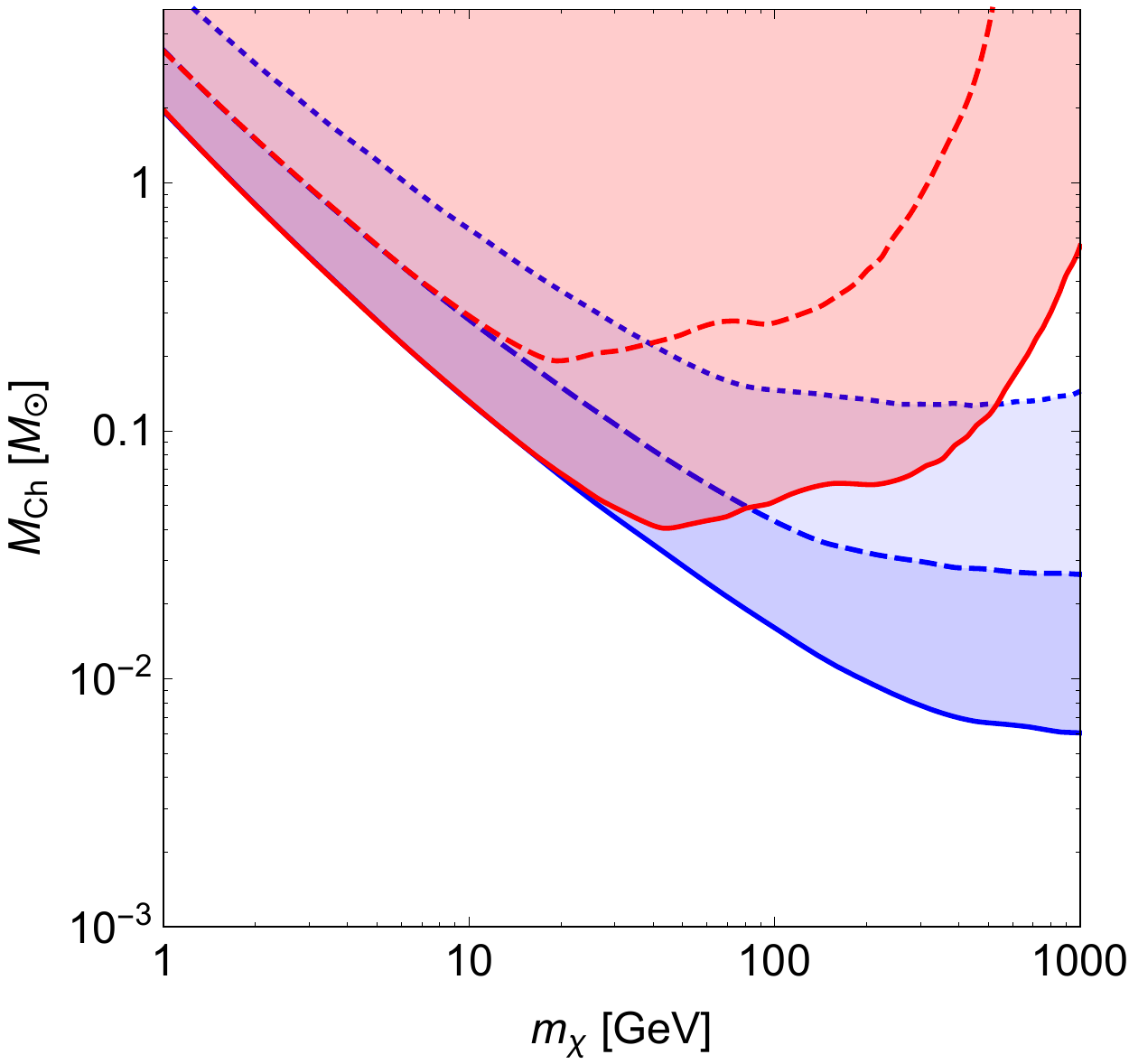} \\
(a) $\alpha = 10^{-2}$ attractive & (b) $\alpha = 10^{-2}$ repulsive \\[6pt]
  \includegraphics[width=.35\textwidth]{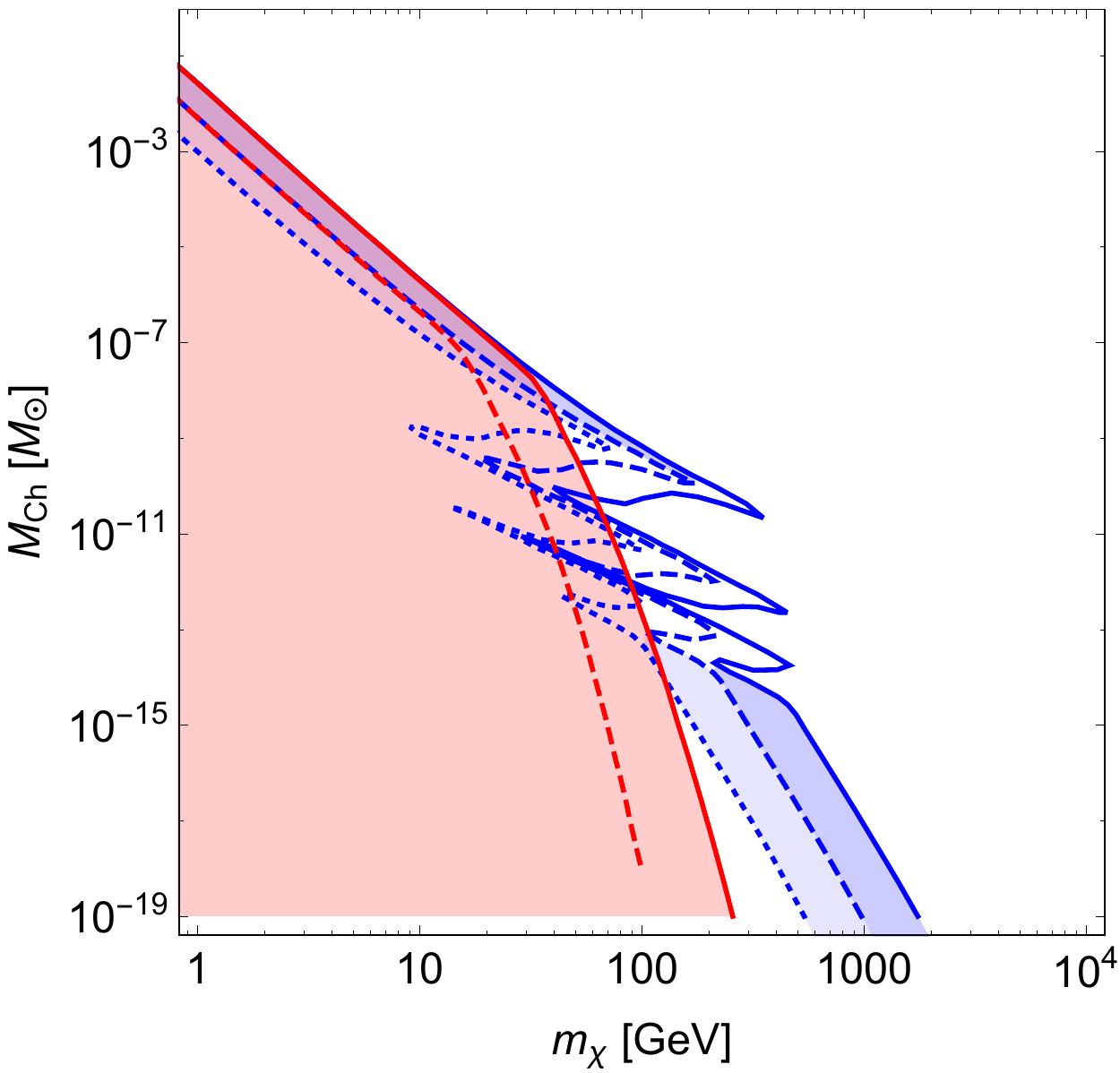} &   \includegraphics[width=.35\textwidth]{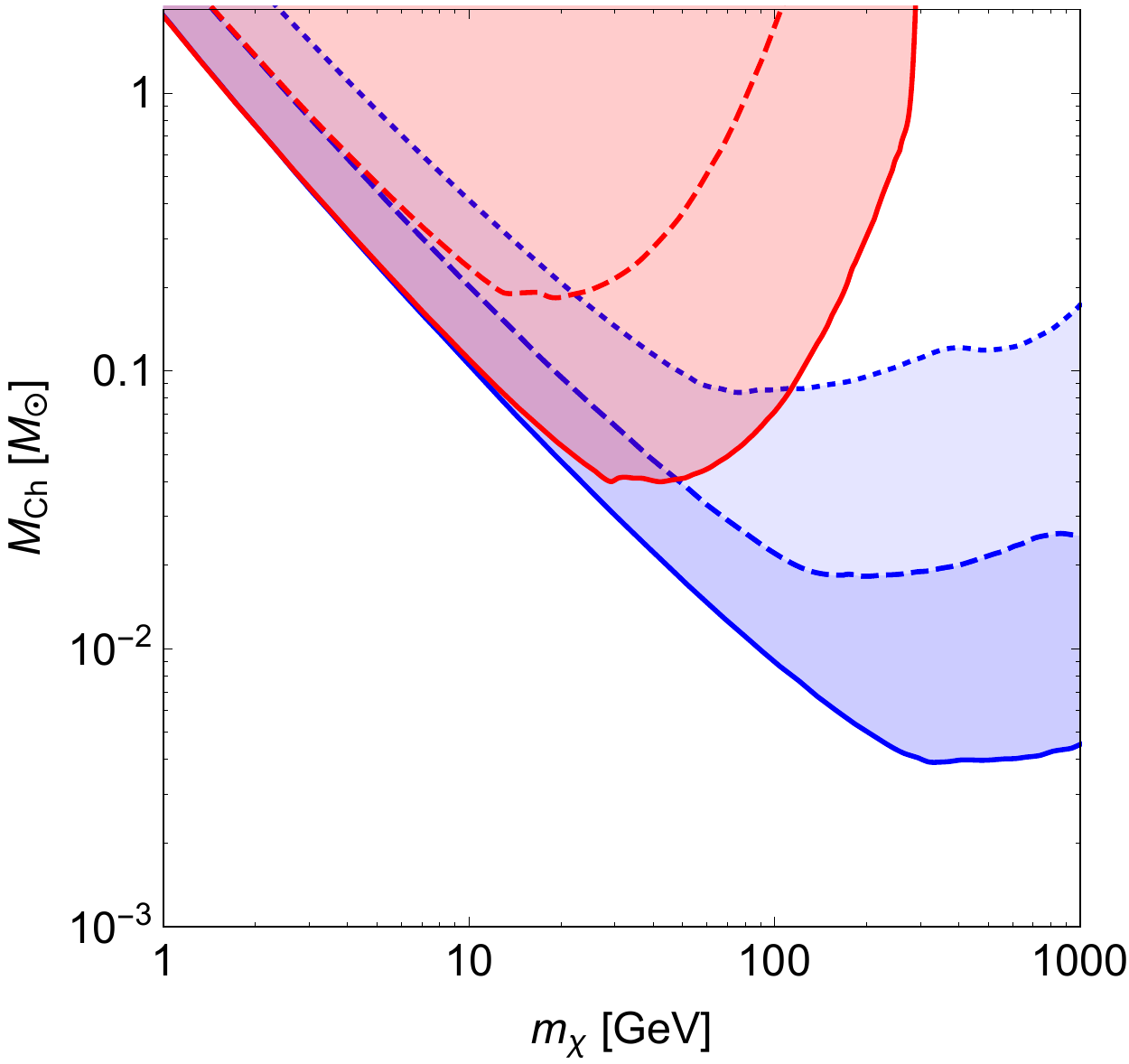} \\
(c) $\alpha = 10^{-3}$ attractive & (d) $\alpha = 10^{-3}$ repulsive \\[6pt]
  \includegraphics[width=.35\textwidth]{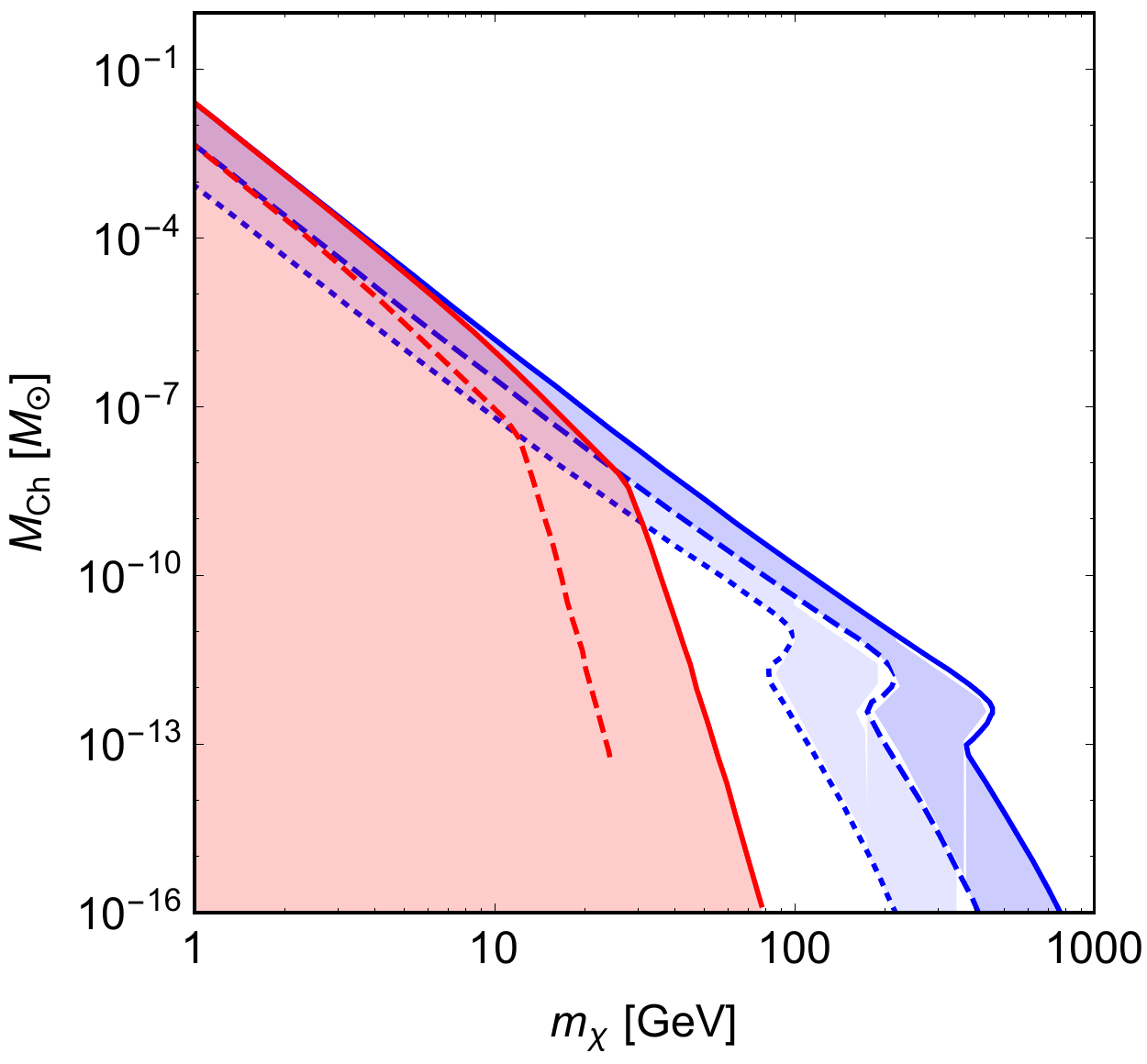} &   \includegraphics[width=.35\textwidth]{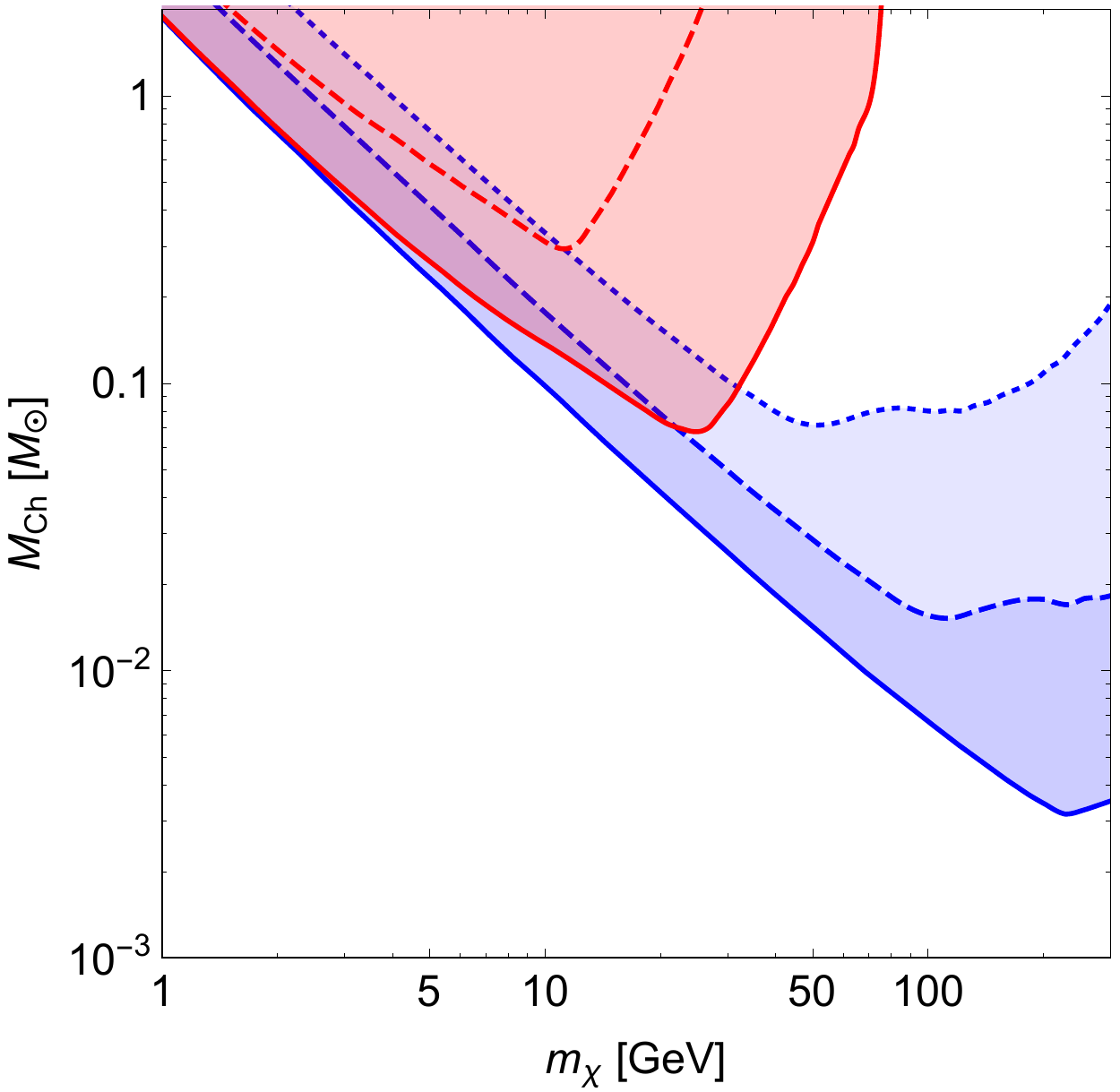} \\
(e) $\alpha = 10^{-4}$ attractive & (f) $\alpha = 10^{-4}$ repulsive
\end{tabular}
\caption{Chandrasekhar mass for dark stars as a function of the DM mass for the parameter space of DM self-interactions shown in Fig.~\ref{fig:parameter}.}
\label{fig:parameter_to_Chandra}
\end{figure*}

 In Fig. \ref{fig:parameter_to_Chandra} we show the maximum mass of dark stars (Chandrasekhar limit) for the DM self-interaction parameter space shown in Fig. \ref{fig:parameter}. A first expected observation is that repulsive self-interactions lead to heavier dark stars compared to attractive ones, since repulsive interactions add to Fermi pressure and therefore more massive configuration can be supported. In fact larger couplings of $\alpha$ (for repulsive interactions) correspond to heavier stars. This can be seen also in Fig. \ref{fig:MvsR2}. On the contrary, larger Yukawa couplings for attractive interactions lead to smaller Chandrasekhar limits. One can see that the difference in the mass of dark stars made of DM particles with attractive and repulsive interactions is large. The mass of the attractive dark stars for the whole parameter space lies below the limits that can be imposed by gravitational lensing. For  repulsive stars, the mass can be a significant fraction of a solar mass, making these stars more visible from the lensing point of view.

\section{Discussion \& Conclusions}

In this paper we studied the possibility that fermionic asymmetric DM can form stars, with self-interactions that can solve well standing problems of the CCDM paradigm. We solved the Tolman-Oppenheimer-Volkoff equation and we derived mass vs radius relations for these stars for both attractive and repulsive DM self-interactions for a range of DM and mediator masses. We also derived upper mass limits for these objects and we studied the hydrostatic stability of the star profiles we examined. 

One issue of fundamental importance that we plan to address in future work is related to the question of how these stars form in the first place.  There are several possibilities where such a scenario can be visualized. One possibility is the creation of high DM density  regions due to adiabatic contraction, caused by baryons~\cite{Blumenthal:1985qy} (see also discussion in~\cite{Gustafsson:2006gr}). Another possibility is the existence of a subdominant strongly interacting DM
component that consists the seed for the gravitational collapse of parts of the DM halo via a gravothermal mechanism~\cite{gravo}. In this case DM self-interactions can lead to a collapse due to the fact that DM-DM collisions can send one of the two particles to a deeper gravitational potential while the second one leaves the area subtracting thus energy from the system that continues until an instability is established. As it has been discussed in~\cite{Pollack:2014rja}, the gravothermal mechanism leads to either gravitational collapse if the DM-DM forces are short range, or to formation of binary systems in case of long range interactions. In our case, the DM-DM interactions are short range because they are mediated by the massive particle $\phi$, (or $\phi_{\mu}$) and therefore the instability leads to formation of DM asymmetric stars. Another possibility is the capture of a significant amount of DM particles by a  supermassive star. After the collapse of the star and the supernova explosion, and since DM particles cannot be blown away in significant amounts by the supernova, a pure DM star or a mixed star with significant amount of baryons can be formed~\cite{Kouvaris:2010vv}. We should emphasize here that all these possibilities do not lead to a total collapse of the whole DM population but rather of a small fraction of it. In view of this, one should not worry for star profiles we present here in the case of repulsive interactions that are within the range of $10^{-7}M_{\odot}-10~M_{\odot}$ constrained by the MACHO~\cite{Alcock:2000ph} and EROS microlensing observations~\cite{Tisserand:2006zx}, simply because these stars do not comprise the whole DM density. 

Another issue we would like to leave out for future work is the discovery signatures for this type of dark stars. Apart from gravitational lensing that can in principle discover objects like this based on the spacetime distortion that their presence can cause, other types of direct signals can exist. If DM communicates with the Standard Model via some portal, e.g. kinetic mixing between the photon and a dark photon, faint photon luminosity should be expected from these stars. Additionally, as we pointed out, for a large range of our parameters, the radius of these stars can be significantly smaller than that of a regular neutron star. This means that asymmetric dark stars can rotate faster than regular neutron stars. Pure dark stars made of DM particles that interact with the Standard Model particles through some portal, or mixed stars can appear as fast rotating pulsars. Rotational frequencies below millisecond are hard to be explained by a regular neutron star. An example of such a case is XTE J1739-285, which allegedly rotates with a frequency of 1122Hz \cite{Kaaret:2006gr}. Any odd looking neutron star is a potential candidate for an asymmetric dark star.

We should mention that for a dark star that has collapsed to a black hole with a mass below $\sim 10^{-19}M_{\odot}$, evaporation of the black hole via Hawking radiation takes place within the age of the universe. Although additional constraints exist~\cite{Capela:2012jz,Pani:2014rca,Capela:2014ita}, there is still a possibility for observing the spectrum of an evaporating black hole in the sky.

Finally we should mention (as one can easily estimate) that in order for a dark star to collide with the earth during earth's lifetime of $t_0\sim4.5 \times 10^9$ years, the dark star mass must be $M< \pi \rho_{\chi} v_0 R_{\oplus}^2 t_0 \simeq 10^{-15}M_{\odot}$ where $\rho_{\chi}=0.3 \text{GeV}/\text{cm}^3$ is the local DM density, $R_{\oplus}$ is the radius of the earth, and $v_0=220\text{km}/\text{sec}$ the velocity dispersion of DM. One can see from Fig.~\ref{fig:parameter_to_Chandra} that $\sim$TeV DM with attractive interactions and a coupling of $\alpha=10^{-4}$ can in fact give dark stars below that mass, making it possible for such a dark star to have collided with the earth in the past. 
\\ \\
This work is supported by the Danish National Research Foundation, Grant No. DNRF90.

%\bibliographystyle{ArXiv}
%\bibliography{nu.bib}

\begin{thebibliography}{99}	


%\cite{Moore:1994yx}
\bibitem{Moore:1994yx} 
  B.~Moore,
  %``Evidence against dissipationless dark matter from observations of galaxy haloes,''
  Nature {\bf 370}, 629 (1994).
  %%CITATION = NATUA,370,629;%%
  %434 citations counted in INSPIRE as of 26 Jun 2015


%\cite{Flores:1994gz}
\bibitem{Flores:1994gz} 
  R.~A.~Flores and J.~R.~Primack,
  %``Observational and theoretical constraints on singular dark matter halos,''
  Astrophys.\ J.\  {\bf 427}, L1 (1994)
  [astro-ph/9402004].
  %%CITATION = ASTRO-PH/9402004;%%
  %373 citations counted in INSPIRE as of 26 Jun 2015


%\cite{Navarro:1996gj}
\bibitem{Navarro:1996gj} 
  J.~F.~Navarro, C.~S.~Frenk and S.~D.~M.~White,
  %``A Universal density profile from hierarchical clustering,''
  Astrophys.\ J.\  {\bf 490}, 493 (1997)
  [astro-ph/9611107].
  %%CITATION = ASTRO-PH/9611107;%%
  %4316 citations counted in INSPIRE as of 26 Jun 2015


%\cite{Klypin:1999uc}
\bibitem{Klypin:1999uc} 
  A.~A.~Klypin, A.~V.~Kravtsov, O.~Valenzuela and F.~Prada,
  %``Where are the missing Galactic satellites?,''
  Astrophys.\ J.\  {\bf 522}, 82 (1999)
  [astro-ph/9901240].
  %%CITATION = ASTRO-PH/9901240;%%
  %1308 citations counted in INSPIRE as of 26 Jun 2015


%\cite{Moore:1999nt}
\bibitem{Moore:1999nt} 
  B.~Moore, S.~Ghigna, F.~Governato, G.~Lake, T.~R.~Quinn, J.~Stadel and P.~Tozzi,
  %``Dark matter substructure within galactic halos,''
  Astrophys.\ J.\  {\bf 524}, L19 (1999)
  [astro-ph/9907411].
  %%CITATION = ASTRO-PH/9907411;%%
  %1483 citations counted in INSPIRE as of 26 Jun 2015


%\cite{Kauffmann:1993gv}
\bibitem{Kauffmann:1993gv} 
  G.~Kauffmann, S.~D.~M.~White and B.~Guiderdoni,
  %``The Formation and Evolution of Galaxies Within Merging Dark Matter Haloes,''
  Mon.\ Not.\ Roy.\ Astron.\ Soc.\  {\bf 264}, 201 (1993).
  %%CITATION = MNRAA,264,201;%%
  %1002 citations counted in INSPIRE as of 26 Jun 2015


%\cite{Liu:2010tn}
\bibitem{Liu:2010tn} 
  L.~Liu, B.~F.~Gerke, R.~H.~Wechsler, P.~S.~Behroozi and M.~T.~Busha,
  %``How Common are the Magellanic Clouds?,''
  Astrophys.\ J.\  {\bf 733}, 62 (2011)
  [arXiv:1011.2255 [astro-ph.CO]].
  %%CITATION = ARXIV:1011.2255;%%
  %49 citations counted in INSPIRE as of 26 Jun 2015


%\cite{Tollerud:2011wt}
\bibitem{Tollerud:2011wt} 
  E.~J.~Tollerud, M.~Boylan-Kolchin, E.~J.~Barton, J.~S.~Bullock and C.~Q.~Trinh,
  %``Small-Scale Structure in the SDSS and LCDM: Isolated L* Galaxies with Bright Satellites,''
  Astrophys.\ J.\  {\bf 738}, 102 (2011)
  [arXiv:1103.1875 [astro-ph.CO]].
  %%CITATION = ARXIV:1103.1875;%%
  %41 citations counted in INSPIRE as of 26 Jun 2015


%\cite{Strigari:2011ps}
\bibitem{Strigari:2011ps} 
  L.~E.~Strigari and R.~H.~Wechsler,
  %``The Cosmic Abundance of Classical Milky Way Satellites,''
  Astrophys.\ J.\  {\bf 749}, 75 (2012)
  [arXiv:1111.2611 [astro-ph.CO]].
  %%CITATION = ARXIV:1111.2611;%%
  %24 citations counted in INSPIRE as of 26 Jun 2015


%\cite{BoylanKolchin:2011de}
\bibitem{BoylanKolchin:2011de} 
  M.~Boylan-Kolchin, J.~S.~Bullock and M.~Kaplinghat,
  %``Too big to fail? The puzzling darkness of massive Milky Way subhaloes,''
  Mon.\ Not.\ Roy.\ Astron.\ Soc.\  {\bf 415}, L40 (2011)
  [arXiv:1103.0007 [astro-ph.CO]].
  %%CITATION = ARXIV:1103.0007;%%
  %256 citations counted in INSPIRE as of 26 Jun 2015


%\cite{Oh:2010mc}
\bibitem{Oh:2010mc} 
  S.~H.~Oh, C.~Brook, F.~Governato, E.~Brinks, L.~Mayer, W.~J.~G.~de Blok, A.~Brooks and F.~Walter,
  %``The central slope of dark matter cores in dwarf galaxies: Simulations vs. THINGS,''
  Astron.\ J.\  {\bf 142}, 24 (2011)
  [arXiv:1011.2777 [astro-ph.CO]].
  %%CITATION = ARXIV:1011.2777;%%
  %71 citations counted in INSPIRE as of 26 Jun 2015


%\cite{Brook:2011nz}
\bibitem{Brook:2011nz} 
  C.~B.~Brook, G.~Stinson, B.~K.~Gibson, R.~Roskar, J.~Wadsley and T.~Quinn,
  %``Hierarchical formation of bulgeless galaxies II: Redistribution of angular momentum via galactic fountains,''
  Mon.\ Not.\ Roy.\ Astron.\ Soc.\  {\bf 419}, 771 (2012)
  [arXiv:1105.2562 [astro-ph.CO]].
  %%CITATION = ARXIV:1105.2562;%%
  %63 citations counted in INSPIRE as of 26 Jun 2015


%\cite{Pontzen:2011ty}
\bibitem{Pontzen:2011ty} 
  A.~Pontzen and F.~Governato,
  %``How supernova feedback turns dark matter cusps into cores,''
  Mon.\ Not.\ Roy.\ Astron.\ Soc.\  {\bf 421}, 3464 (2012)
  [arXiv:1106.0499 [astro-ph.CO]].
  %%CITATION = ARXIV:1106.0499;%%
  %197 citations counted in INSPIRE as of 26 Jun 2015


%\cite{Governato:2012fa}
\bibitem{Governato:2012fa} 
  F.~Governato, A.~Zolotov, A.~Pontzen, C.~Christensen, S.~H.~Oh, A.~M.~Brooks, T.~Quinn and S.~Shen {\it et al.},
  %``Cuspy No More: How Outflows Affect the Central Dark Matter and Baryon Distribution in Lambda CDM Galaxies,''
  Mon.\ Not.\ Roy.\ Astron.\ Soc.\  {\bf 422}, 1231 (2012)
  [arXiv:1202.0554 [astro-ph.CO]].
  %%CITATION = ARXIV:1202.0554;%%
  %183 citations counted in INSPIRE as of 26 Jun 2015


%\cite{Vogelsberger:2012ku}
\bibitem{Vogelsberger:2012ku} 
  M.~Vogelsberger, J.~Zavala and A.~Loeb,
  %``Subhaloes in Self-Interacting Galactic Dark Matter Haloes,''
  Mon.\ Not.\ Roy.\ Astron.\ Soc.\  {\bf 423}, 3740 (2012)
  [arXiv:1201.5892 [astro-ph.CO]].
  %%CITATION = ARXIV:1201.5892;%%
  %120 citations counted in INSPIRE as of 26 Jun 2015


%\cite{Rocha:2012jg}
\bibitem{Rocha:2012jg} 
  M.~Rocha, A.~H.~G.~Peter, J.~S.~Bullock, M.~Kaplinghat, S.~Garrison-Kimmel, J.~Onorbe and L.~A.~Moustakas,
  %``Cosmological Simulations with Self-Interacting Dark Matter I: Constant Density Cores and Substructure,''
  Mon.\ Not.\ Roy.\ Astron.\ Soc.\  {\bf 430}, 81 (2013)
  [arXiv:1208.3025 [astro-ph.CO]].
  %%CITATION = ARXIV:1208.3025;%%
  %143 citations counted in INSPIRE as of 26 Jun 2015


%\cite{Zavala:2012us}
\bibitem{Zavala:2012us} 
  J.~Zavala, M.~Vogelsberger and M.~G.~Walker,
  %``Constraining Self-Interacting Dark Matter with the Milky Way's dwarf spheroidals,''
  Monthly Notices of the Royal Astronomical Society: Letters {\bf 431}, L20 (2013)
  [arXiv:1211.6426 [astro-ph.CO]].
  %%CITATION = ARXIV:1211.6426;%%
  %56 citations counted in INSPIRE as of 26 Jun 2015


%\cite{Peter:2012jh}
\bibitem{Peter:2012jh} 
  A.~H.~G.~Peter, M.~Rocha, J.~S.~Bullock and M.~Kaplinghat,
  %``Cosmological Simulations with Self-Interacting Dark Matter II: Halo Shapes vs. Observations,''
  Mon.\ Not.\ Roy.\ Astron.\ Soc.\  {\bf 430}, 105 (2013)
  [arXiv:1208.3026 [astro-ph.CO]].
  %%CITATION = ARXIV:1208.3026;%%
  %115 citations counted in INSPIRE as of 26 Jun 2015


%\cite{Spergel:1999mh}
\bibitem{Spergel:1999mh} 
  D.~N.~Spergel and P.~J.~Steinhardt,
  %``Observational evidence for selfinteracting cold dark matter,''
  Phys.\ Rev.\ Lett.\  {\bf 84}, 3760 (2000)
  [astro-ph/9909386].
  %%CITATION = ASTRO-PH/9909386;%%
  %747 citations counted in INSPIRE as of 26 Jun 2015


%\cite{Wandelt:2000ad}
\bibitem{Wandelt:2000ad} 
  B.~D.~Wandelt, R.~Dave, G.~R.~Farrar, P.~C.~McGuire, D.~N.~Spergel and P.~J.~Steinhardt,
  %``Selfinteracting dark matter,''
  astro-ph/0006344.
  %%CITATION = ASTRO-PH/0006344;%%
  %116 citations counted in INSPIRE as of 26 Jun 2015


%\cite{Faraggi:2000pv}
\bibitem{Faraggi:2000pv} 
  A.~E.~Faraggi and M.~Pospelov,
  %``Selfinteracting dark matter from the hidden heterotic string sector,''
  Astropart.\ Phys.\  {\bf 16}, 451 (2002)
  [hep-ph/0008223].
  %%CITATION = HEP-PH/0008223;%%
  %28 citations counted in INSPIRE as of 26 Jun 2015


%\cite{Mohapatra:2001sx}
\bibitem{Mohapatra:2001sx} 
  R.~N.~Mohapatra, S.~Nussinov and V.~L.~Teplitz,
  %``Mirror matter as selfinteracting dark matter,''
  Phys.\ Rev.\ D {\bf 66}, 063002 (2002)
  [hep-ph/0111381].
  %%CITATION = HEP-PH/0111381;%%
  %54 citations counted in INSPIRE as of 26 Jun 2015


%\cite{Kusenko:2001vu}
\bibitem{Kusenko:2001vu} 
  A.~Kusenko and P.~J.~Steinhardt,
  %``Q ball candidates for selfinteracting dark matter,''
  Phys.\ Rev.\ Lett.\  {\bf 87}, 141301 (2001)
  [astro-ph/0106008].
  %%CITATION = ASTRO-PH/0106008;%%
  %69 citations counted in INSPIRE as of 26 Jun 2015


%\cite{Loeb:2010gj}
\bibitem{Loeb:2010gj} 
  A.~Loeb and N.~Weiner,
  %``Cores in Dwarf Galaxies from Dark Matter with a Yukawa Potential,''
  Phys.\ Rev.\ Lett.\  {\bf 106}, 171302 (2011)
  [arXiv:1011.6374 [astro-ph.CO]].
  %%CITATION = ARXIV:1011.6374;%%
  %112 citations counted in INSPIRE as of 26 Jun 2015


%\cite{Kouvaris:2011gb}
\bibitem{Kouvaris:2011gb} 
  C.~Kouvaris,
  %``Limits on Self-Interacting Dark Matter,''
  Phys.\ Rev.\ Lett.\  {\bf 108}, 191301 (2012)
  [arXiv:1111.4364 [astro-ph.CO]].
  %%CITATION = ARXIV:1111.4364;%%
  %41 citations counted in INSPIRE as of 26 Jun 2015


%\cite{Vogelsberger:2012sa}
\bibitem{Vogelsberger:2012sa} 
  M.~Vogelsberger and J.~Zavala,
  %``Direct detection of self-interacting dark matter,''
  Mon.\ Not.\ Roy.\ Astron.\ Soc.\  {\bf 430}, 1722 (2013)
  [arXiv:1211.1377 [astro-ph.CO]].
  %%CITATION = ARXIV:1211.1377;%%
  %23 citations counted in INSPIRE as of 26 Jun 2015


%\cite{Tulin:2013teo}
\bibitem{Tulin:2013teo} 
  S.~Tulin, H.~B.~Yu and K.~M.~Zurek,
  %``Beyond Collisionless Dark Matter: Particle Physics Dynamics for Dark Matter Halo Structure,''
  Phys.\ Rev.\ D {\bf 87}, no. 11, 115007 (2013)
  [arXiv:1302.3898 [hep-ph]].
  %%CITATION = ARXIV:1302.3898;%%
  %94 citations counted in INSPIRE as of 26 Jun 2015


%\cite{Kaplinghat:2013xca}
\bibitem{Kaplinghat:2013xca} 
  M.~Kaplinghat, R.~E.~Keeley, T.~Linden and H.~B.~Yu,
  %``Tying Dark Matter to Baryons with Self-interactions,''
  Phys.\ Rev.\ Lett.\  {\bf 113}, 021302 (2014)
  [arXiv:1311.6524 [astro-ph.CO]].
  %%CITATION = ARXIV:1311.6524;%%
  %19 citations counted in INSPIRE as of 26 Jun 2015


%\cite{Kaplinghat:2013yxa}
\bibitem{Kaplinghat:2013yxa} 
  M.~Kaplinghat, S.~Tulin and H.~B.~Yu,
  %``Direct Detection Portals for Self-interacting Dark Matter,''
  Phys.\ Rev.\ D {\bf 89}, no. 3, 035009 (2014)
  [arXiv:1310.7945 [hep-ph]].
  %%CITATION = ARXIV:1310.7945;%%
  %22 citations counted in INSPIRE as of 26 Jun 2015


%\cite{Cline:2013pca}
\bibitem{Cline:2013pca} 
  J.~M.~Cline, Z.~Liu, G.~Moore and W.~Xue,
  %``Scattering properties of dark atoms and molecules,''
  Phys.\ Rev.\ D {\bf 89}, no. 4, 043514 (2014)
  [arXiv:1311.6468 [hep-ph]].
  %%CITATION = ARXIV:1311.6468;%%
  %21 citations counted in INSPIRE as of 26 Jun 2015


%\cite{Cline:2013zca}
\bibitem{Cline:2013zca} 
  J.~M.~Cline, Z.~Liu, G.~Moore and W.~Xue,
  %``Composite strongly interacting dark matter,''
  Phys.\ Rev.\ D {\bf 90}, no. 1, 015023 (2014)
  [arXiv:1312.3325 [hep-ph]].
  %%CITATION = ARXIV:1312.3325;%%
  %32 citations counted in INSPIRE as of 26 Jun 2015


%\cite{Petraki:2014uza}
\bibitem{Petraki:2014uza} 
  K.~Petraki, L.~Pearce and A.~Kusenko,
  %``Self-interacting asymmetric dark matter coupled to a light massive dark photon,''
  JCAP {\bf 1407}, 039 (2014)
  [arXiv:1403.1077 [hep-ph]].
  %%CITATION = ARXIV:1403.1077;%%
  %19 citations counted in INSPIRE as of 26 Jun 2015


%\cite{Buckley:2014hja}
\bibitem{Buckley:2014hja} 
  M.~R.~Buckley, J.~Zavala, F.~Y.~Cyr-Racine, K.~Sigurdson and M.~Vogelsberger,
  %``Scattering, Damping, and Acoustic Oscillations: Simulating the Structure of Dark Matter Halos with Relativistic Force Carriers,''
  Phys.\ Rev.\ D {\bf 90}, no. 4, 043524 (2014)
  [arXiv:1405.2075 [astro-ph.CO]].
  %%CITATION = ARXIV:1405.2075;%%
  %18 citations counted in INSPIRE as of 26 Jun 2015


%\cite{Boddy:2014yra}
\bibitem{Boddy:2014yra} 
  K.~K.~Boddy, J.~L.~Feng, M.~Kaplinghat and T.~M.~P.~Tait,
  %``Self-Interacting Dark Matter from a Non-Abelian Hidden Sector,''
  Phys.\ Rev.\ D {\bf 89}, no. 11, 115017 (2014)
  [arXiv:1402.3629 [hep-ph]].
  %%CITATION = ARXIV:1402.3629;%%
  %30 citations counted in INSPIRE as of 26 Jun 2015


%\cite{Schutz:2014nka}
\bibitem{Schutz:2014nka} 
  K.~Schutz and T.~R.~Slatyer,
  %``Self-Scattering for Dark Matter with an Excited State,''
  JCAP {\bf 1501}, no. 01, 021 (2015)
  [arXiv:1409.2867 [hep-ph]].
  %%CITATION = ARXIV:1409.2867;%%
  %8 citations counted in INSPIRE as of 26 Jun 2015


%\cite{Feng:2009mn}
\bibitem{Feng:2009mn} 
  J.~L.~Feng, M.~Kaplinghat, H.~Tu and H.~B.~Yu,
  %``Hidden Charged Dark Matter,''
  JCAP {\bf 0907}, 004 (2009)
  [arXiv:0905.3039 [hep-ph]].
  %%CITATION = ARXIV:0905.3039;%%
  %142 citations counted in INSPIRE as of 26 Jun 2015


%\cite{Feng:2009hw}
\bibitem{Feng:2009hw} 
  J.~L.~Feng, M.~Kaplinghat and H.~B.~Yu,
  %``Halo Shape and Relic Density Exclusions of Sommerfeld-Enhanced Dark Matter Explanations of Cosmic Ray Excesses,''
  Phys.\ Rev.\ Lett.\  {\bf 104}, 151301 (2010)
  [arXiv:0911.0422 [hep-ph]].
  %%CITATION = ARXIV:0911.0422;%%
  %124 citations counted in INSPIRE as of 26 Jun 2015


%\cite{Markevitch:2003at}
\bibitem{Markevitch:2003at} 
  M.~Markevitch, A.~H.~Gonzalez, D.~Clowe, A.~Vikhlinin, L.~David, W.~Forman, C.~Jones and S.~Murray {\it et al.},
  %``Direct constraints on the dark matter self-interaction cross-section from the merging galaxy cluster 1E0657-56,''
  Astrophys.\ J.\  {\bf 606}, 819 (2004)
  [astro-ph/0309303].
  %%CITATION = ASTRO-PH/0309303;%%
  %229 citations counted in INSPIRE as of 26 Jun 2015


%\cite{Burgess:2000yq}
\bibitem{Burgess:2000yq} 
  C.~P.~Burgess, M.~Pospelov and T.~ter Veldhuis,
  %``The Minimal model of nonbaryonic dark matter: A Singlet scalar,''
  Nucl.\ Phys.\ B {\bf 619}, 709 (2001)
  [hep-ph/0011335].
  %%CITATION = HEP-PH/0011335;%%
  %469 citations counted in INSPIRE as of 26 Jun 2015


%\cite{Patt:2006fw}
\bibitem{Patt:2006fw} 
  B.~Patt and F.~Wilczek,
  %``Higgs-field portal into hidden sectors,''
  hep-ph/0605188.
  %%CITATION = HEP-PH/0605188;%%
  %310 citations counted in INSPIRE as of 26 Jun 2015


%\cite{Andreas:2008xy}
\bibitem{Andreas:2008xy} 
  S.~Andreas, T.~Hambye and M.~H.~G.~Tytgat,
  %``WIMP dark matter, Higgs exchange and DAMA,''
  JCAP {\bf 0810}, 034 (2008)
  [arXiv:0808.0255 [hep-ph]].
  %%CITATION = ARXIV:0808.0255;%%
  %152 citations counted in INSPIRE as of 26 Jun 2015


%\cite{Andreas:2010dz}
\bibitem{Andreas:2010dz} 
  S.~Andreas, C.~Arina, T.~Hambye, F.~S.~Ling and M.~H.~G.~Tytgat,
  %``A light scalar WIMP through the Higgs portal and CoGeNT,''
  Phys.\ Rev.\ D {\bf 82}, 043522 (2010)
  [arXiv:1003.2595 [hep-ph]].
  %%CITATION = ARXIV:1003.2595;%%
  %129 citations counted in INSPIRE as of 26 Jun 2015


%\cite{Djouadi:2011aa}
\bibitem{Djouadi:2011aa} 
  A.~Djouadi, O.~Lebedev, Y.~Mambrini and J.~Quevillon,
  %``Implications of LHC searches for Higgs--portal dark matter,''
  Phys.\ Lett.\ B {\bf 709}, 65 (2012)
  [arXiv:1112.3299 [hep-ph]].
  %%CITATION = ARXIV:1112.3299;%%
  %169 citations counted in INSPIRE as of 26 Jun 2015


%\cite{Pospelov:2011yp}
\bibitem{Pospelov:2011yp} 
  M.~Pospelov and A.~Ritz,
  %``Higgs decays to dark matter: beyond the minimal model,''
  Phys.\ Rev.\ D {\bf 84}, 113001 (2011)
  [arXiv:1109.4872 [hep-ph]].
  %%CITATION = ARXIV:1109.4872;%%
  %48 citations counted in INSPIRE as of 26 Jun 2015


%\cite{Greljo:2013wja}
\bibitem{Greljo:2013wja} 
  A.~Greljo, J.~Julio, J.~F.~Kamenik, C.~Smith and J.~Zupan,
  %``Constraining Higgs mediated dark matter interactions,''
  JHEP {\bf 1311}, 190 (2013)
  [arXiv:1309.3561 [hep-ph]].
  %%CITATION = ARXIV:1309.3561;%%
  %23 citations counted in INSPIRE as of 26 Jun 2015


%\cite{Bhattacherjee:2013jca}
\bibitem{Bhattacherjee:2013jca} 
  B.~Bhattacherjee, S.~Matsumoto, S.~Mukhopadhyay and M.~M.~Nojiri,
  %``Phenomenology of light fermionic asymmetric dark matter,''
  JHEP {\bf 1310}, 032 (2013)
  [arXiv:1306.5878 [hep-ph]].
  %%CITATION = ARXIV:1306.5878;%%
  %9 citations counted in INSPIRE as of 26 Jun 2015


%\cite{Kouvaris:2014uoa}
\bibitem{Kouvaris:2014uoa} 
  C.~Kouvaris, I.~M.~Shoemaker and K.~Tuominen,
  %``Self-Interacting Dark Matter through the Higgs Portal,''
  Phys.\ Rev.\ D {\bf 91}, no. 4, 043519 (2015)
  [arXiv:1411.3730 [hep-ph]].
  %%CITATION = ARXIV:1411.3730;%%
  %7 citations counted in INSPIRE as of 26 Jun 2015


%\cite{Alcock:2000ph}
\bibitem{Alcock:2000ph} 
  C.~Alcock {\it et al.}  [MACHO Collaboration],
  %``The MACHO project: Microlensing results from 5.7 years of LMC observations,''
  Astrophys.\ J.\  {\bf 542}, 281 (2000)
  [astro-ph/0001272].
  %%CITATION = ASTRO-PH/0001272;%%
  %481 citations counted in INSPIRE as of 26 Jun 2015


%\cite{Tisserand:2006zx}
\bibitem{Tisserand:2006zx} 
  P.~Tisserand {\it et al.}  [EROS-2 Collaboration],
  %``Limits on the Macho Content of the Galactic Halo from the EROS-2 Survey of the Magellanic Clouds,''
  Astron.\ Astrophys.\  {\bf 469}, 387 (2007)
  [astro-ph/0607207].
  %%CITATION = ASTRO-PH/0607207;%%
  %167 citations counted in INSPIRE as of 26 Jun 2015


%\cite{Spolyar:2007qv}
\bibitem{Spolyar:2007qv} 
  D.~Spolyar, K.~Freese and P.~Gondolo,
  %``Dark matter and the first stars: a new phase of stellar evolution,''
  Phys.\ Rev.\ Lett.\  {\bf 100}, 051101 (2008)
  [arXiv:0705.0521 [astro-ph]].
  %%CITATION = ARXIV:0705.0521;%%
  %119 citations counted in INSPIRE as of 26 Jun 2015


%\cite{Freese:2008hb}
\bibitem{Freese:2008hb} 
  K.~Freese, P.~Gondolo, J.~A.~Sellwood and D.~Spolyar,
  %``Dark Matter Densities during the Formation of the First Stars and in Dark Stars,''
  Astrophys.\ J.\  {\bf 693}, 1563 (2009)
  [arXiv:0805.3540 [astro-ph]].
  %%CITATION = ARXIV:0805.3540;%%
  %45 citations counted in INSPIRE as of 26 Jun 2015


%\cite{Freese:2008wh}
\bibitem{Freese:2008wh} 
  K.~Freese, P.~Bodenheimer, D.~Spolyar and P.~Gondolo,
  %``Stellar Structure of Dark Stars: a first phase of Stellar Evolution due to Dark Matter Annihilation,''
  Astrophys.\ J.\  {\bf 685}, L101 (2008)
  [arXiv:0806.0617 [astro-ph]].
  %%CITATION = ARXIV:0806.0617;%%
  %70 citations counted in INSPIRE as of 26 Jun 2015


%\cite{Leung:2011zz}
\bibitem{Leung:2011zz} 
  S.~C.~Leung, M.~C.~Chu and L.~M.~Lin,
  %``Dark-matter admixed neutron stars,''
  Phys.\ Rev.\ D {\bf 84}, 107301 (2011)
  [arXiv:1111.1787 [astro-ph.CO]].
  %%CITATION = ARXIV:1111.1787;%%
  %8 citations counted in INSPIRE as of 26 Jun 2015


%\cite{Leung:2013pra}
\bibitem{Leung:2013pra} 
  S.-C.~Leung, M.-C.~Chu, L.-M.~Lin and K.-W.~Wong,
  %``Dark-matter admixed white dwarfs,''
  Phys.\ Rev.\ D {\bf 87}, no. 12, 123506 (2013)
  [arXiv:1305.6142 [astro-ph.CO]].
  %%CITATION = ARXIV:1305.6142;%%
  %8 citations counted in INSPIRE as of 26 Jun 2015


%\cite{Pollack:2014rja}
\bibitem{Pollack:2014rja} 
  J.~Pollack, D.~N.~Spergel and P.~J.~Steinhardt,
  %``Supermassive Black Holes from Ultra-Strongly Self-Interacting Dark Matter,''
  Astrophys.\ J.\  {\bf 804}, no. 2, 131 (2015)
  [arXiv:1501.00017 [astro-ph.CO]].
  %%CITATION = ARXIV:1501.00017;%%
  %2 citations counted in INSPIRE as of 26 Jun 2015


%\cite{Nussinov:1985xr}
\bibitem{Nussinov:1985xr} 
  S.~Nussinov,
  %``Technocosmology: Could A Technibaryon Excess Provide A 'natural' Missing Mass Candidate?,''
  Phys.\ Lett.\ B {\bf 165}, 55 (1985).
  %%CITATION = PHLTA,B165,55;%%
  %265 citations counted in INSPIRE as of 26 Jun 2015


%\cite{Barr:1990ca}
\bibitem{Barr:1990ca} 
  S.~M.~Barr, R.~S.~Chivukula and E.~Farhi,
  %``Electroweak Fermion Number Violation and the Production of Stable Particles in the Early Universe,''
  Phys.\ Lett.\ B {\bf 241}, 387 (1990).
  %%CITATION = PHLTA,B241,387;%%
  %211 citations counted in INSPIRE as of 26 Jun 2015


%\cite{Gudnason:2006yj}
\bibitem{Gudnason:2006yj} 
  S.~B.~Gudnason, C.~Kouvaris and F.~Sannino,
  %``Dark Matter from new Technicolor Theories,''
  Phys.\ Rev.\ D {\bf 74}, 095008 (2006)
  [hep-ph/0608055].
  %%CITATION = HEP-PH/0608055;%%
  %162 citations counted in INSPIRE as of 26 Jun 2015


%\cite{Foadi:2008qv}
\bibitem{Foadi:2008qv} 
  R.~Foadi, M.~T.~Frandsen and F.~Sannino,
  %``Technicolor Dark Matter,''
  Phys.\ Rev.\ D {\bf 80}, 037702 (2009)
  [arXiv:0812.3406 [hep-ph]].
  %%CITATION = ARXIV:0812.3406;%%
  %84 citations counted in INSPIRE as of 26 Jun 2015


%\cite{Dietrich:2006cm}
\bibitem{Dietrich:2006cm} 
  D.~D.~Dietrich and F.~Sannino,
  %``Conformal window of SU(N) gauge theories with fermions in higher dimensional representations,''
  Phys.\ Rev.\ D {\bf 75}, 085018 (2007)
  [hep-ph/0611341].
  %%CITATION = HEP-PH/0611341;%%
  %260 citations counted in INSPIRE as of 26 Jun 2015


%\cite{Sannino:2009za}
\bibitem{Sannino:2009za} 
  F.~Sannino,
  %``Conformal Dynamics for TeV Physics and Cosmology,''
  Acta Phys.\ Polon.\ B {\bf 40}, 3533 (2009)
  [arXiv:0911.0931 [hep-ph]].
  %%CITATION = ARXIV:0911.0931;%%
  %193 citations counted in INSPIRE as of 26 Jun 2015


%\cite{Ryttov:2008xe}
\bibitem{Ryttov:2008xe} 
  T.~A.~Ryttov and F.~Sannino,
  %``Ultra Minimal Technicolor and its Dark Matter TIMP,''
  Phys.\ Rev.\ D {\bf 78}, 115010 (2008)
  [arXiv:0809.0713 [hep-ph]].
  %%CITATION = ARXIV:0809.0713;%%
  %108 citations counted in INSPIRE as of 26 Jun 2015


%\cite{Sannino:2008nv}
\bibitem{Sannino:2008nv} 
  F.~Sannino and R.~Zwicky,
  %``Unparticle and Higgs as Composites,''
  Phys.\ Rev.\ D {\bf 79}, 015016 (2009)
  [arXiv:0810.2686 [hep-ph]].
  %%CITATION = ARXIV:0810.2686;%%
  %67 citations counted in INSPIRE as of 26 Jun 2015


%\cite{Kaplan:2009ag}
\bibitem{Kaplan:2009ag} 
  D.~E.~Kaplan, M.~A.~Luty and K.~M.~Zurek,
  %``Asymmetric Dark Matter,''
  Phys.\ Rev.\ D {\bf 79}, 115016 (2009)
  [arXiv:0901.4117 [hep-ph]].
  %%CITATION = ARXIV:0901.4117;%%
  %314 citations counted in INSPIRE as of 26 Jun 2015


%\cite{Frandsen:2009mi}
\bibitem{Frandsen:2009mi} 
  M.~T.~Frandsen and F.~Sannino,
  %``iTIMP: isotriplet Technicolor Interacting Massive Particle as Dark Matter,''
  Phys.\ Rev.\ D {\bf 81}, 097704 (2010)
  [arXiv:0911.1570 [hep-ph]].
  %%CITATION = ARXIV:0911.1570;%%
  %52 citations counted in INSPIRE as of 26 Jun 2015


%\cite{MarchRussell:2011fi}
\bibitem{MarchRussell:2011fi} 
  J.~March-Russell and M.~McCullough,
  %``Asymmetric Dark Matter via Spontaneous Co-Genesis,''
  JCAP {\bf 1203}, 019 (2012)
  [arXiv:1106.4319 [hep-ph]].
  %%CITATION = ARXIV:1106.4319;%%
  %58 citations counted in INSPIRE as of 26 Jun 2015


%\cite{Frandsen:2011cg}
\bibitem{Frandsen:2011cg} 
  M.~T.~Frandsen, F.~Kahlhoefer, S.~Sarkar and K.~Schmidt-Hoberg,
  %``Direct detection of dark matter in models with a light Z',''
  JHEP {\bf 1109}, 128 (2011)
  [arXiv:1107.2118 [hep-ph]].
  %%CITATION = ARXIV:1107.2118;%%
  %58 citations counted in INSPIRE as of 26 Jun 2015


%\cite{Gao:2011ka}
\bibitem{Gao:2011ka} 
  X.~Gao, Z.~Kang and T.~Li,
  %``Origins of the Isospin Violation of Dark Matter Interactions,''
  JCAP {\bf 1301}, 021 (2013)
  [arXiv:1107.3529 [hep-ph]].
  %%CITATION = ARXIV:1107.3529;%%
  %38 citations counted in INSPIRE as of 26 Jun 2015


%\cite{Arina:2011cu}
\bibitem{Arina:2011cu} 
  C.~Arina and N.~Sahu,
  %``Asymmetric Inelastic Inert Doublet Dark Matter from Triplet Scalar Leptogenesis,''
  Nucl.\ Phys.\ B {\bf 854}, 666 (2012)
  [arXiv:1108.3967 [hep-ph]].
  %%CITATION = ARXIV:1108.3967;%%
  %50 citations counted in INSPIRE as of 26 Jun 2015


%\cite{Buckley:2011ye}
\bibitem{Buckley:2011ye} 
  M.~R.~Buckley and S.~Profumo,
  %``Regenerating a Symmetry in Asymmetric Dark Matter,''
  Phys.\ Rev.\ Lett.\  {\bf 108}, 011301 (2012)
  [arXiv:1109.2164 [hep-ph]].
  %%CITATION = ARXIV:1109.2164;%%
  %46 citations counted in INSPIRE as of 26 Jun 2015


%\cite{Lewis:2011zb}
\bibitem{Lewis:2011zb} 
  R.~Lewis, C.~Pica and F.~Sannino,
  %``Light Asymmetric Dark Matter on the Lattice: SU(2) Technicolor with Two Fundamental Flavors,''
  Phys.\ Rev.\ D {\bf 85}, 014504 (2012)
  [arXiv:1109.3513 [hep-ph]].
  %%CITATION = ARXIV:1109.3513;%%
  %53 citations counted in INSPIRE as of 26 Jun 2015


%\cite{Davoudiasl:2011fj}
\bibitem{Davoudiasl:2011fj} 
  H.~Davoudiasl, D.~E.~Morrissey, K.~Sigurdson and S.~Tulin,
  %``Baryon Destruction by Asymmetric Dark Matter,''
  Phys.\ Rev.\ D {\bf 84}, 096008 (2011)
  [arXiv:1106.4320 [hep-ph]].
  %%CITATION = ARXIV:1106.4320;%%
  %61 citations counted in INSPIRE as of 26 Jun 2015


%\cite{Graesser:2011wi}
\bibitem{Graesser:2011wi} 
  M.~L.~Graesser, I.~M.~Shoemaker and L.~Vecchi,
  %``Asymmetric WIMP dark matter,''
  JHEP {\bf 1110}, 110 (2011)
  [arXiv:1103.2771 [hep-ph]].
  %%CITATION = ARXIV:1103.2771;%%
  %94 citations counted in INSPIRE as of 26 Jun 2015


%\cite{Bell:2011tn}
\bibitem{Bell:2011tn} 
  N.~F.~Bell, K.~Petraki, I.~M.~Shoemaker and R.~R.~Volkas,
  %``Pangenesis in a Baryon-Symmetric Universe: Dark and Visible Matter via the Affleck-Dine Mechanism,''
  Phys.\ Rev.\ D {\bf 84}, 123505 (2011)
  [arXiv:1105.3730 [hep-ph]].
  %%CITATION = ARXIV:1105.3730;%%
  %59 citations counted in INSPIRE as of 26 Jun 2015


%\cite{Cheung:2011if}
\bibitem{Cheung:2011if} 
  C.~Cheung and K.~M.~Zurek,
  %``Affleck-Dine Cogenesis,''
  Phys.\ Rev.\ D {\bf 84}, 035007 (2011)
  [arXiv:1105.4612 [hep-ph]].
  %%CITATION = ARXIV:1105.4612;%%
  %62 citations counted in INSPIRE as of 26 Jun 2015


%\cite{Petraki:2013wwa}
\bibitem{Petraki:2013wwa} 
  K.~Petraki and R.~R.~Volkas,
  %``Review of asymmetric dark matter,''
  Int.\ J.\ Mod.\ Phys.\ A {\bf 28}, 1330028 (2013)
  [arXiv:1305.4939 [hep-ph]].
  %%CITATION = ARXIV:1305.4939;%%
  %103 citations counted in INSPIRE as of 26 Jun 2015


%\cite{Zurek:2013wia}
\bibitem{Zurek:2013wia} 
  K.~M.~Zurek,
  %``Asymmetric Dark Matter: Theories, Signatures, and Constraints,''
  Phys.\ Rept.\  {\bf 537}, 91 (2014)
  [arXiv:1308.0338 [hep-ph]].
  %%CITATION = ARXIV:1308.0338;%%
  %107 citations counted in INSPIRE as of 26 Jun 2015


%\cite{Shapiro:1983du}
\bibitem{Shapiro:1983du} 
  S.~L.~Shapiro and S.~A.~Teukolsky,
  %``Black holes, white dwarfs, and neutron stars: The physics of compact objects,''
  New York, USA: Wiley (1983) 645 p
  %13 citations counted in INSPIRE as of 26 Jun 2015

\bibitem{Weinberg:1972} 
  S.~Weinberg,
  %``Gravitation and Cosmology: Principles and applications of the general theory of relativity,''
  New York, USA: Wiley (1972) 657 p
  %13 citations counted in INSPIRE as of 26 Jun 2015


%\cite{Finkbeiner:2010sm}
\bibitem{Finkbeiner:2010sm} 
  D.~P.~Finkbeiner, L.~Goodenough, T.~R.~Slatyer, M.~Vogelsberger and N.~Weiner,
  %``Consistent Scenarios for Cosmic-Ray Excesses from Sommerfeld-Enhanced Dark Matter Annihilation,''
  JCAP {\bf 1105}, 002 (2011)
  [arXiv:1011.3082 [hep-ph]].
  %%CITATION = ARXIV:1011.3082;%%
  %57 citations counted in INSPIRE as of 26 Jun 2015


\bibitem{Yukawaderivation}
  N.~Gauthier,
  %``Yukawa potential approach to the nuclear binding energy formula''
Am.\ J.\ Phys.\ {\bf 58}, 375 (1990)
  doi: 10.1119/1.16176

%\cite{Blumenthal:1985qy}
\bibitem{Blumenthal:1985qy} 
  G.~R.~Blumenthal, S.~M.~Faber, R.~Flores and J.~R.~Primack,
  %``Contraction of Dark Matter Galactic Halos Due to Baryonic Infall,''
  Astrophys.\ J.\  {\bf 301}, 27 (1986).
  %%CITATION = ASJOA,301,27;%%
  %704 citations counted in INSPIRE as of 26 Jun 2015


%\cite{Gustafsson:2006gr}
\bibitem{Gustafsson:2006gr} 
  M.~Gustafsson, M.~Fairbairn and J.~Sommer-Larsen,
  %``Baryonic Pinching of Galactic Dark Matter Haloes,''
  Phys.\ Rev.\ D {\bf 74}, 123522 (2006)
  [astro-ph/0608634].
  %%CITATION = ASTRO-PH/0608634;%%
  %95 citations counted in INSPIRE as of 26 Jun 2015


\bibitem{gravo}
  D. Lynden-Bell and P. P. Eggleton, 
  %"On the consequences of the gravothermal catastrophe,"
MNRAS 191 (May, 1980) 483-498.

%\cite{Kouvaris:2010vv}
\bibitem{Kouvaris:2010vv} 
  C.~Kouvaris and P.~Tinyakov,
  %``Can Neutron stars constrain Dark Matter?,''
  Phys.\ Rev.\ D {\bf 82}, 063531 (2010)
  [arXiv:1004.0586 [astro-ph.GA]].
  %%CITATION = ARXIV:1004.0586;%%
  %47 citations counted in INSPIRE as of 26 Jun 2015


%\cite{Kaaret:2006gr}
\bibitem{Kaaret:2006gr} 
  P.~Kaaret, Z.~Prieskorn, J.~J.~M.~i.~'t Zand, S.~Brandt, N.~Lund, S.~Mereghetti, D.~Gotz and E.~Kuulkers {\it et al.},
  %``Discovery of 1122-Hz X-Ray Burst Oscillations from the Neutron-Star X-Ray Transient XTE J1739-285,''
  Astrophys.\ J.\  {\bf 657}, L97 (2007)
  [astro-ph/0611716].
  %%CITATION = ASTRO-PH/0611716;%%
  %74 citations counted in INSPIRE as of 26 Jun 2015


%\cite{Capela:2012jz}
\bibitem{Capela:2012jz} 
  F.~Capela, M.~Pshirkov and P.~Tinyakov,
  %``Constraints on Primordial Black Holes as Dark Matter Candidates from Star Formation,''
  Phys.\ Rev.\ D {\bf 87}, no. 2, 023507 (2013)
  [arXiv:1209.6021 [astro-ph.CO]].
  %%CITATION = ARXIV:1209.6021;%%
  %16 citations counted in INSPIRE as of 26 Jun 2015


%\cite{Pani:2014rca}
\bibitem{Pani:2014rca} 
  P.~Pani and A.~Loeb,
  %``Tidal capture of a primordial black hole by a neutron star: implications for constraints on dark matter,''
  JCAP {\bf 1406}, 026 (2014)
  [arXiv:1401.3025 [astro-ph.CO]].
  %%CITATION = ARXIV:1401.3025;%%
  %12 citations counted in INSPIRE as of 26 Jun 2015


%\cite{Capela:2014ita}
\bibitem{Capela:2014ita} 
  F.~Capela, M.~Pshirkov and P.~Tinyakov,
  %``Adiabatic contraction revisited: implications for primordial black holes,''
  Phys.\ Rev.\ D {\bf 90}, no. 8, 083507 (2014)
  [arXiv:1403.7098 [astro-ph.CO]].
  %%CITATION = ARXIV:1403.7098;%%
  %6 citations counted in INSPIRE as of 26 Jun 2015




\end{thebibliography}

\end{document}